\documentclass[traditabstract]{aa} 
\usepackage[lofdepth,lotdepth]{subfig}
\usepackage{graphicx}
\usepackage{wrapfig}
\usepackage[varg]{txfonts}
\usepackage{longtable}
\usepackage{lscape}
\usepackage{natbib}
\usepackage{url}
\usepackage{color}
\usepackage{txfonts}
\usepackage{float}
\usepackage{color}
\usepackage{multirow}
\usepackage{caption}
\usepackage{appendix}
\usepackage{threeparttable}

\bibpunct{(}{)}{;}{a}{}{,} 

\newcommand\teff{T_{\rm eff}}

\begin{document}
   \title{The origin and evolution of the odd-Z iron-peak elements \\
   Sc, V, Mn, and Co in the Milky Way stellar disk\thanks{This paper includes data gathered with the 6.5 meter Magellan Telescopes located at the Las Campanas Observatory, Chile; the Nordic Optical Telescope (NOT) on La Palma, Spain; the Very Large Telescope (VLT) at the European Southern Observatory (ESO) on Paranal, Chile (ESO Proposal ID 69.B-0277 and 72.B-0179); the ESO 1.5-m, 2.2-m. and 3.6-m telescopes on La Silla, Chile (ESO Proposal ID 65.L-0019, 67.B-0108, 76.B-0416, 82.B-0610); and data from UVES Paranal Observatory Project (ESO DDT Program ID 266.D-5655).}\fnmsep\thanks{Full versions of Tables~2 and 5 are only available at the CDS via anonymous ftp to \tt{cdsarc.u-strasbg.fr (130.79.128.5)} or via \tt{http://cdsarc.u-strasbg.fr/viz-bin/qcat?J/A+A/XXX/XXX}}}
   
     \titlerunning{Sc, V, Mn, and Co in the Milky Way stellar disk}
     \authorrunning{Battistini \& Bensby}


   \author{Chiara Battistini
          \and
         Thomas Bensby }

   \institute{Lund Observatory, Department of Astronomy and Theoretical Physics, Box 43, SE-221\,00, Lund, Sweden
              \\
              \email{chiara@astro.lu.se}
             }

   \date{Received 13 November 2014 / Accepted 3 February 2015}
 
 \abstract
 {Elements heavier than Li are produced in the interiors of stars. However, for many elements the exact production sites and the timescales on which they are dispersed into the interstellar medium are unknown. Having a clear picture on the origins of the elements is important for our ability to trace and understand the formation and chemical evolution of the Milky Way and its stellar populations. 
  }
 {The aim of this study is to investigate the origin and evolution of Sc, V, Mn, and Co for a homogeneous and statistically significant sample of stars probing the different populations of the Milky Way, in particular the thin and thick disks.
 }
 {Using high-resolution spectra obtained with the MIKE, FEROS, SOFIN, FIES, UVES, and HARPS spectrographs, we determine Sc, V, Mn, and Co abundances for a large sample of F and G dwarfs in the solar neighborhood. The method is based on spectral synthesis and using one-dimensional, plane-parallel, local thermodynamic equilibrium (LTE) model stellar atmospheres calculated with the MARCS 2012 code. The non-LTE (NLTE) corrections from the literature were applied to Mn and Co.
 }
 {We find that the abundance trends derived for Sc (594 stars), V (466 stars), and Co (567 stars) are very similar to what has been observed for the $\alpha$-elements in the thin and thick disks. On the contrary, Mn (569 stars) is generally underabundant relative to the Sun (i.e., $\rm [Mn/Fe]<0$) for $\rm [Fe/H] < 0$. In addition, for Mn, when NLTE corrections are applied, the trend changes and is almost flat over the entire metallicity range of the stars in our sample  ($\rm -2 \lesssim [Fe/H] \lesssim +0.4$). The [Sc/Fe]-[Fe/H] abundance trends show a small separation between the thin and thick disks, while for V and Co they completely overlap. For Mn there is a small difference in [Mn/Fe], but only when NLTE corrections are used. Comparisons with Ti as a reference element show flat trends for all the elements except for Mn that show well separated [Mn/Ti]-[Ti/H] trends for the thin and thick disks.
 }
 {The elements Sc and V present trends compatible with production from type II supernovae (SNII) events. In addition, Sc clearly shows a metallicity dependence for [Fe/H] < $-$1. Instead, Mn  is produced in SNII events for $\rm [Fe/H] \lesssim -0.4$ and then type Ia supernovae start to produce Mn. Finally, Co appears to be produced mainly in SNII with suggestion of enrichment from hypernovae at low metallicities.
 }
  



   \maketitle
%

\section{Introduction}
All elements heavier than Li are produced via nuclear processes in the interiors of stars. The timescale for the production of elements is strongly related to the mass of the stars, because not all stars can produce the same elements during their lifetime \citep{Burbidge1957}. Massive stars ($\gtrsim$\,8\,M$_{\odot}$) end their lives exploding as  type II supernovae (SNII), releasing all the elements produced during their entire life cycles and during the explosions into the surrounding interstellar medium (ISM). Stars ending their life in this way are the main source for enrichment of the $\alpha$-elements (e.g., Si, Ti, Ca, Mg) created through the summing up of He nuclei. The elements in the iron group (like Sc and Co) are also partly produced by SN\,II. The enrichment of these elements is expected to happen on a short timescale because the lifetime of the massive stars that end their lives as SNII is   a few million  years \citep{Matteucci1986}.  Low mass stars can also pollute the ISM through  type Ia supernovae (SNIa), releasing significant quantities of Fe in the ISM but none or very little of the $\alpha$-elements (e.g., \citealt{Thielemann2002}). This process requires more time because the binary system needs to evolve away from the main sequence, meaning that the SNIa will be delayed relative to SNII (e.g., \citealt{Matteucci2009}). Depending on the different scenario involved and depending on other parameters, like star formation rate or the mass of the hosting galaxy, the delay for SNIa with respect to SNII ranges from 0.3 to 1\,Gyr (e.g., \citealt{Valiante2009}).

The iron-peak group includes elements in the periodic table from Sc to Ge. Even if they are grouped in one broad category they are produced in complex nucleosynthesis processes that result in an enrichment of the ISM not always following Fe, meaning that the sites of production of these elements are often different than for Fe. The focus of this study is on the odd-Z iron-peak elements,  Sc, V, Mn, and Co. The analysis of these elements needs to be done carefully, taking into account that they suffer from hyperfine structure (hfs) that splits the absorption lines of odd-Z elements into multiple components. The production sites for the elements are uncertain and  the stellar yields of these elements is also under debate. For instance, Sc shares some behavior with the $\alpha$-elements and is predicted to be produced during carbon and neon burning phases in massive stars \citep{Woosley1995} while V, Mn, and Co are believed to be mainly produced in explosive silicon burning in SNII \citep{Woosley1995} and to a smaller extent in SNIa \citep{Bravo2012}.  In addition, models of yields do not always  agree with observations because of the scarse knowledge of certain processes in stellar evolution but also because the production of elements in supernovae explosions are highly sensitive to explosion parameters \citep{Kobayashi2006,Kobayashi2011}. Another factor is that the presence of any dependence with the metallicity of the SNIa progenitors or the rate of occurrence of SNIa are unclear \citep{Kobayashi1998}. In addition, chemodynamical models for certain elements do not always  agree with observations. 
The Galaxy is composed of different stellar populations and these stellar populations show different abundance trends for several chemical elements. This is especially clear for the $\alpha$-elements, which show a distinct separation between the thin and thick disks (e.g., \citealt{Bensby2003,Bensby2005,Bensby2014, Reddy2003,Reddy2006,Adibekyan2012}). The combination of well-defined stellar populations and elements with well-constrained formation sites as reference elements, can help us to study the origin of other elements with less constrained production sites, such as  the odd-Z iron-peak elements Sc, V, Mn, and Co.

Main sequence stars have lifetimes that are comparable to the age of the Galaxy and they preserve mainly unchanged atmospheres keeping the information about the chemical composition of their birth clouds \citep{Lambert1989,Freeman2002}. These are the ideal objects with which to trace the origin of the elements and to try to reconstruct the history of the Milky Way. In this study we are using a large sample of  F and G dwarfs in the solar neighborhood to try to characterize possible differences in abundances trends in the Galactic thin and thick disks and the origin of the odd-Z iron-peak elements, Sc, V, Mn, and Co. \\

The paper is organized as follow. In Sect.~\ref{sect:abundance_analysis} the stellar sample together with the abundance analysis are described. Section~\ref{sect:results} shows the results for Sc, V, Mn, and Co. Finally, Sects.~\ref{sect:origin_Mn} and~\ref{sect:origin_other} describe the possible origins for odd-Z iron-peak, while Sect.~\ref{sect:summary} summarizes our findings and provides a future outlook.

\begin{table*}\tiny
\begin{threeparttable}[b]
\caption{Lines used and determined solar abundances.\label{tab:lines}\tablefootmark{$^\dagger$}}
\setlength{\tabcolsep}{1mm}
\centering
\begin{tabular}{c c c c c c c c c c c | c}
\hline
\hline
\\
Name & $\lambda$ & EP & FEROS & UVES & MIKE 1 & MIKE 2 & MIKE 3 & MIKE 4 & <MIKE> & $\sigma$ <MIKE> & Solar abundance\\
 & $[\AA]$ & eV &  R = 48000 & R = 80000  & R = 65000 & R = 65000 & R = 65000 & R = 42000 & & & from  \cite{Grevesse2007}\\
 \\
 \hline
 \\
\ion{Sc}{ii} & 5526 & 1.768 & 2.97 & 3.00 & 3.01 & 2.99 & 3.03 & 3.07 & 3.03 & 0.03 & 3.17$\pm0.1$\\
\ion{Sc}{ii} & 5657 & 1.507 & 3.06 & 3.12 & 3.16 & 3.14 & 3.14 & 3.19 & 3.16 & 0.02 &\\
\ion{Sc}{ii} & 5667 & 1.500 & 3.10 & 3.12 & 3.14 & 3.15 & 3.12 & 3.12 & 3.13 & 0.02 &\\
\ion{Sc}{ii} & 5684 & 1.507 & 2.84 & 2.86 & 2.88 & 2.89 & 2.88 & 2.89 & 2.89 & 0.01 &\\
\ion{Sc}{ii} & 6245 & 1.507 & 3.12 & 3.13 & 3.12 & 3.12 & 3.12 & 3.17 & 3.13 & 0.03 &\\
\\
\hline
\\
\ion{V}{i} & 5670 & 1.081 & 3.90 & 3.89 & 3.92 & 3.93 & 3.92 & 3.96 & 3.93 & 0.02 & 4.00$\pm$0.02\\
\ion{V}{i} & 6081 & 1.051 & 3.86 & 3.88 & 3.88 & 3.86 & 3.92 & 3.88 & 3.89 & 0.03 &\\
\ion{V}{i} & 6251 & 0.287 & 3.86 & 3.88 & 3.91 & 3.87 & 3.92 & 3.88 & 3.90 & 0.02 &\\
\ion{V}{i} & 6274 & 0.267 & 3.94 & 3.92 & 3.93 & 3.91 & 3.89 & 3.88 & 3.91 & 0.02 &\\
\ion{V}{i} & 6285 & 0.280 & 3.90 & 3.92 & 3.89 & 3.86 & 3.88 & 3.84 & 3.87 & 0.02 &\\
\\
\hline
\\
\ion{Mn}{i} & 5394 & 0.000 & 5.18 & 5.20 & 5.20 & 5.19 & 5.20 & 5.20 & 5.20 & 0.01 & 5.39$\pm$0.03\\
\ion{Mn}{i} & 5432 & 0.000 & 5.30 & 5.28 & 5.32 & 5.31 & 5.32 & 5.32 & 5.32 & 0.01 &\\
\ion{Mn}{i} & 6013 & 3.073 & 5.25 & 5.27 & 5.31 & 5.30 & 5.30 & 5.34 & 5.31 & 0.02 &\\
\ion{Mn}{i} & 6016 & 3.073 & 5.38 & 5.40 & 5.45 & 5.45 & 5.46 & 5.50 & 5.47 & 0.03 &\\
\\
\hline
\\
\ion{Co}{i} & 5301 & 1.711 & 4.90 & 4.91 & 4.95 & 4.94 & 4.94 & 4.96 & 4.95 & 0.01 & 4.92$\pm$0.08\\
\ion{Co}{i} & 5342 & 4.020 & 4.81 & 4.85 & 4.88 & 4.88 & 4.88 & 4.89 & 4.88 & 0.01 &\\
\ion{Co}{i} & 5352 & 3.577 & 4.76 & 4.74 & 4.79 & 4.80 & 4.79 & 4.81 & 4.80 & 0.01 &\\
\ion{Co}{i} & 5647 & 2.280 & 4.88 & 4.87 & 4.88 & 4.84 & 4.87 & 4.88 & 4.87 & 0.02 &\\
\\
\hline
\end{tabular}
\tablefoot{\tablefoottext{$\dagger$}{Columns 2 and 3 contain atomic data for the lines (element and ionization stage and the classic energy potential respectively). Columns 4-9 list the single abundances for the different instruments and resolutions. Columns 10 and 11 show the median abundance values for MIKE spectra for the different observation runs and the one $\sigma$ dispersion, respectively. The mean abundances are used as solar abundance for MIKE spectra observed in the run with $R = 55\,000$ (see Table 1 in \cite{Bensby2014}) since no solar spectra is available in this setting. For HARPS spectra the mean values from all the abundances derived in the solar spectra are used. In Col. 12 are listed the solar abundances values from \cite{Grevesse2007}.}}
\end{threeparttable}
\end{table*}

\begin{table*}[t]\tiny
\caption{Abundance from single lines.\label{tab:abundances_single_lines}\tablefootmark{$^\dagger$}}
\setlength{\tabcolsep}{1 mm}
\centering
\begin{tabular}{c | c c c c c | c c c c c | c c c c | c c c c }
\hline
\multirow{2}{*} {HIP}& \multicolumn{5}{c|} {$\mathrm{\epsilon(Sc)}$} & \multicolumn{5}{c|}{$\mathrm{\epsilon(V)}$} & \multicolumn{4}{c|}{$\mathrm{\epsilon(Mn})$} & \multicolumn{4}{c}{$\mathrm{\epsilon(Co)}$}\\
\cline{2-19}
 & 5526 & 5657 & 5667 & 5684 & 6245 & 5670 & 6081 & 6251 & 6274 & 6285 & 5394 & 5432 & 6013 & 6016 & 5301 & 5342 & 5352 & 5647\\
\hline
\\
80 & 3.21 & 3.35 & $--$ & $--$ & 3.32 & $--$ & $--$ & $--$ & $--$ & $--$ & $--$ & $--$ & $--$ & $--$ & 4.43 & 4.40 & 4.44 & $--$\\
305 & 3.31 & 3.37 & $--$ & $--$ & $--$ & $--$ & 4.05 & 4.11 & $--$ & 4.09 & 5.34 & 5.41 & 5.46 & 5.59 & 4.52 & $--$ & 4.98 & $--$\\
407 & 3.10 & 3.25 & $--$ & 2.95 & $--$ & 4.06 & $--$ & $--$ & 4.06 & $--$ & 5.31 & 5.41 & 5.48 & 5.61 & 5.07 & 4.96 & 4.91 & 4.94\\
\vdots & \vdots & \vdots & \vdots & \vdots & \vdots & \vdots & \vdots & \vdots & \vdots & \vdots & \vdots & \vdots & \vdots & \vdots & \vdots & \vdots & \vdots & \vdots \\
\\
\hline
\end{tabular}
\tablefoot{\tablefoottext{$\dagger$}{Abundances from the single line analyzed in this work. The table is only available in electronic form at the CDS}}
\end{table*}

\section{Abundance analysis \label{sect:abundance_analysis}}
\subsection{Stellar sample}
The sample consists of the 714 F and G dwarf stars from \cite{Bensby2014} selected to probe the abundance structure of the Milky Way stellar disk. High-resolution and high signal-to-noise spectra were obtained during several observation runs in the years 2000-2007 using the high-resolution spectrographs SOFIN and FIES on NOT, UVES on VLT, HARPS on the ESO 3.6 m telescope on La Silla, FEROS on the ESO 1.52 m telescope on La Silla, and MIKE on Magellan. The resolving power of the spectra is between $R = 45000-120000$, and the signal-to-noise is generally greater than $S/N \gtrsim 200$. The observations and data reduction are described in \cite{Bensby2014}.

\begin{figure}[ht]
\centering
\resizebox{0.9\hsize}{!}{
\includegraphics[bb=0 70 792 600,clip]{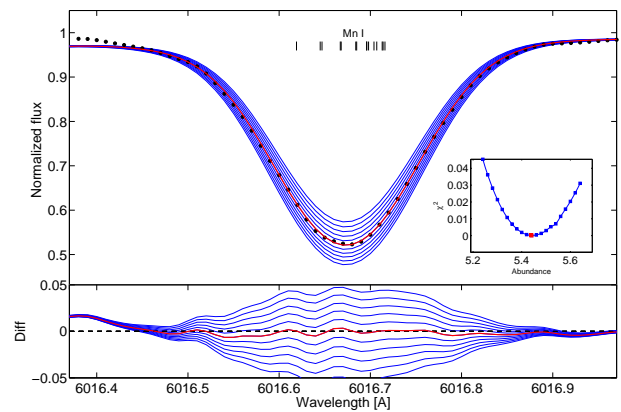}}
\caption{Example of a well-fitted line in the case of solar spectrum for the \ion{Mn}{i} line at 6016\,{\AA}. The upper panel shows eleven synthetic spectra with different abundances in steps of 0.04\,dex (blue lines). The positions of the hyperfine components are indicated at the top. The difference between the observed spectrum and the synthetic spectra is plotted in the lower panel with the best abundance spectrum plotted as a red line in both the upper and lower panel. The inset presents the $\chi^{2}$ fit to determine the best abundance value expressed as $\log \epsilon_{X}$ (shown as a red dot); the 22 blue dots are the $\chi^{2}$ values for all the spectra created by SME.\label{fig:good_fit}}
\end{figure}


\subsection{Stellar parameters and abundances for other elements}

Stellar parameters, ages, and elemental abundances for O, Na, Mg, Al, Si, Ca, Ti, Cr, Fe, Ni, Zn, Y, and Ba were determined for all 714 stars in \cite{Bensby2014}. Briefly, the determination of stellar parameters was performed by requiring excitation balance of abundances from \ion{Fe}{i} lines for the effective temperature ($\teff$), ionization balance between \ion{Fe}{i} and  \ion{Fe}{ii} lines for surface gravity ($\log g$), and that abundances from \ion{Fe}{i} lines be independent of reduced line strength to get microturbulence parameter ($\mathrm{\xi_{t}}$). The ionization and excitation balance method utilised in \cite{Bensby2014} and widely in other works (e.g., \citealt{Adibekyan2012}) results in an apparently flat lower main sequence (see Figs.~9 and 12 in \citealt{Bensby2014}). The reason is not clear, and small empirical corrections were applied to $\teff$ and $\log g$ (see \cite{Bensby2014} for details). 

Stellar ages were determined from the $\alpha$-enhanced Yonsei-Yale isochrones by \cite{Demarque2004}  using probability distribution functions as explained in \cite{Bensby2014}.
\subsection{Line synthesis of Sc, V, Mn, Co}
\subsubsection{Line lists and atomic data}
Since the iron peak elements analyzed in this study are odd-Z elements, this means that hyperfine splitting is important and must be taken into account in the creation of synthetic spectra. The hfs effect makes spectral lines broader with higher equivalent widths than if considered as single lines and for strong lines the result is to desaturate them \citep{Prochaska2000}. For the analysis of Mn we use the same four lines as in \cite{Feltzing2007} while for Sc, V, and Co, the lines were selected from \cite{Prochaska2000}.
Wavelengths and $\log(gf)$ for the hfs components were also taken from \cite{Prochaska2000}. Blending lines and other lines close to our lines of interest were taken from Vienna Atomic Line Database (VALD, \citealt{Piskunov1995,Ryabchikova1997,Kupka1999,Kupka2000}). The lines that we use are listed in Table~\ref{tab:lines}. The full line lists, including hfs components and other nearby lines taken from VALD are given in Tables~\ref{tab:hfs1}-~\ref{tab:hfs4} in Appendix A.


\subsubsection{Synthetic spectra creation}\label{sect:syn_creation}

The  abundance analysis is based on fitting the synthetic spectra to the observed ones using a standard local thermodynamic equilibrium (LTE) analysis with 1D plane-parallel model atmospheres calculated with the MARCS 2012 code \citep{Gustafsson2008}. The creation of synthetic spectra was done with SME (Spectroscopy Made Easy, \citealt{Valenti1996,Valenti2005}), and even though SME is a powerful tool with which to derive stellar parameters and elemental abundances, we only used it as a spectral synthesizer with fixed stellar parameters from \cite{Bensby2014} and abundances passed from the user.

\subsubsection{Abundance determination}\label{sect:abun_det}
To determine  the abundances several synthetic spectra were created with different abundances in wide steps (0.25\,dex) to find a first best value through minimizing an unnormalized $\chi^{2} $, defined as
\\
\begin{equation}
\begin{centering}
\chi^{2} = \sum{\frac{(observed - synthetic)^{2}}{observed}}
\end{centering}
.\end{equation}Another 21 synthetic spectra were then generated for each line with different abundance values spanning a range of $\pm 0.2$\,dex around the first best value in steps of 0.02\,dex. Finally, the best abundance value for each line was found using another unnormalized $\chi^{2}$. An example of the result of this procedure is shown in Fig.~\ref{fig:good_fit} for one \ion{Mn}{i} line of one of the available solar spectrum. More examples of the different synthetic spectra compared with the observed ones for the other lines are shown in Figs.~\ref{fig:solar_spectrum1}--\ref{fig:solar_spectrum4} in Appendix B for the solar spectrum.

\begin{figure*}[ht]
\centering
\resizebox{1\hsize}{!}{
\includegraphics[bb=20 15 792 300,clip]{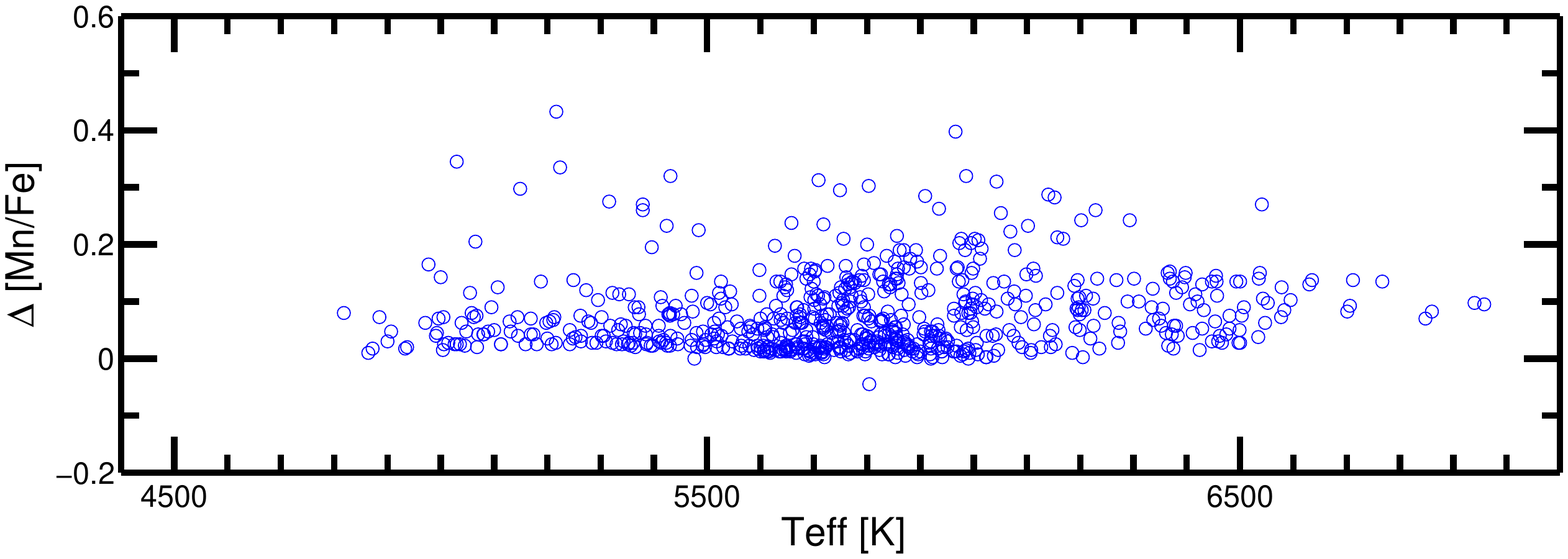}
\includegraphics[bb=20 15 792 300,clip]{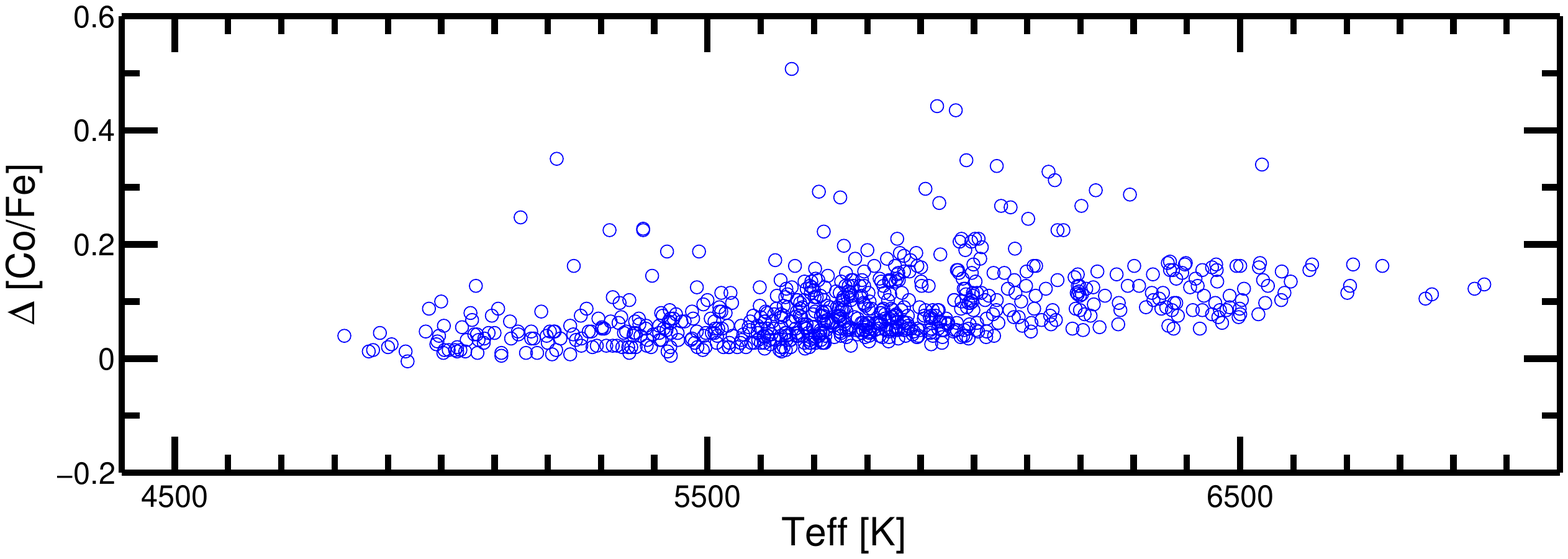}}
\resizebox{1\hsize}{!}{
\includegraphics[bb=20 15 792 290,clip]{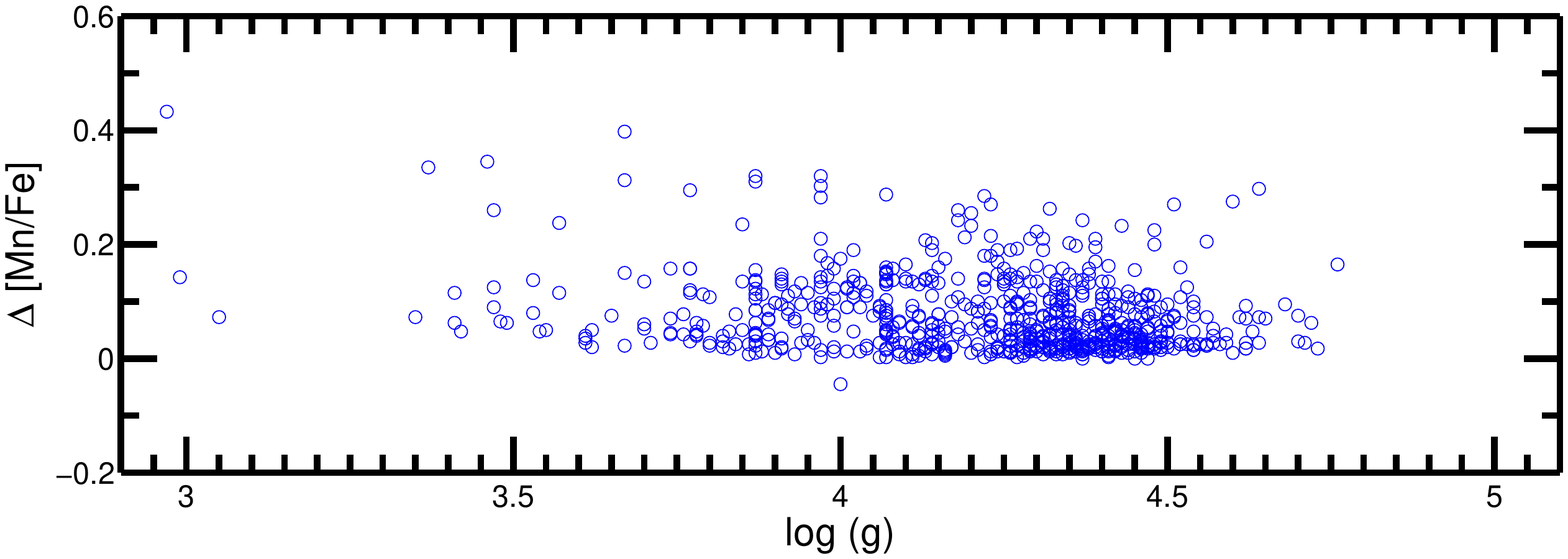}
\includegraphics[bb=20 15 792 290,clip]{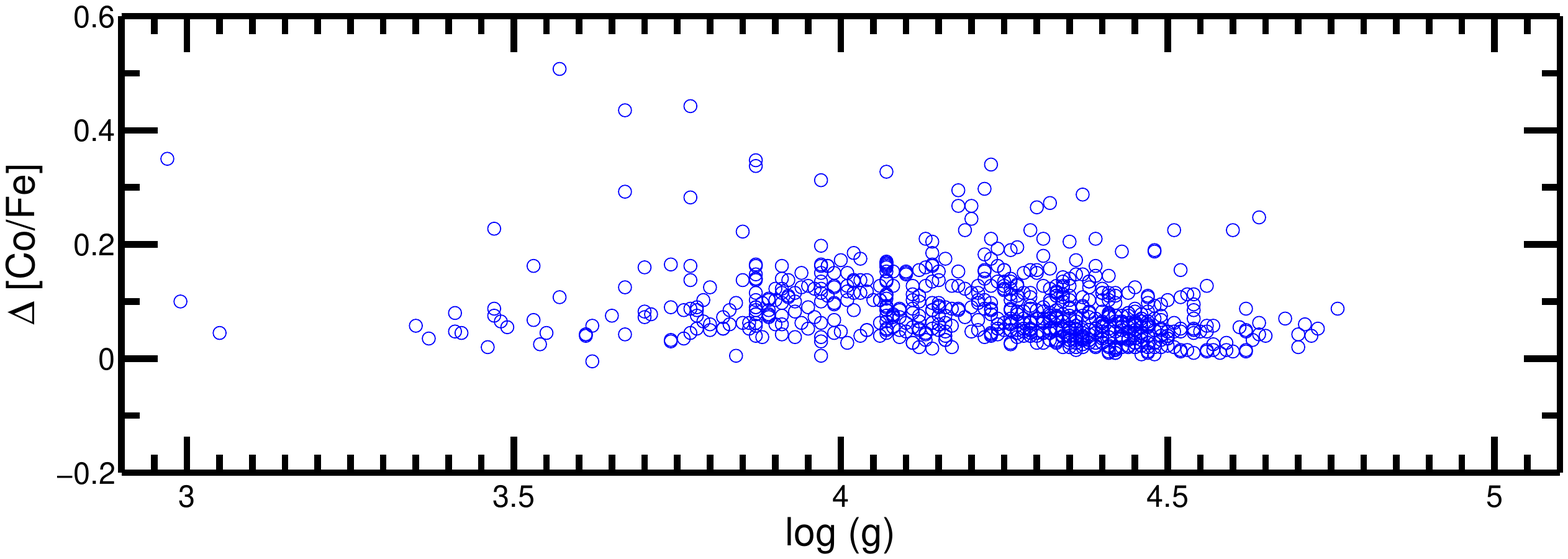}}
\resizebox{1\hsize}{!}{
\includegraphics[bb=20 30 792 290,clip]{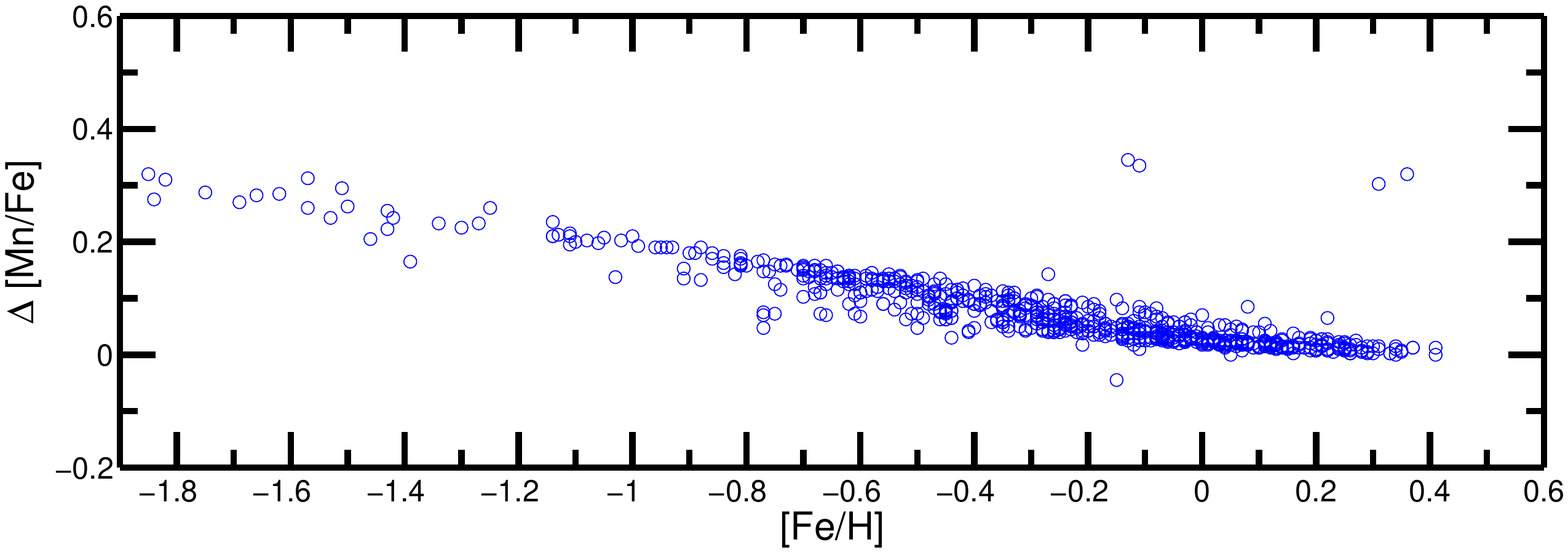}
\includegraphics[bb=20 30 792 290,clip]{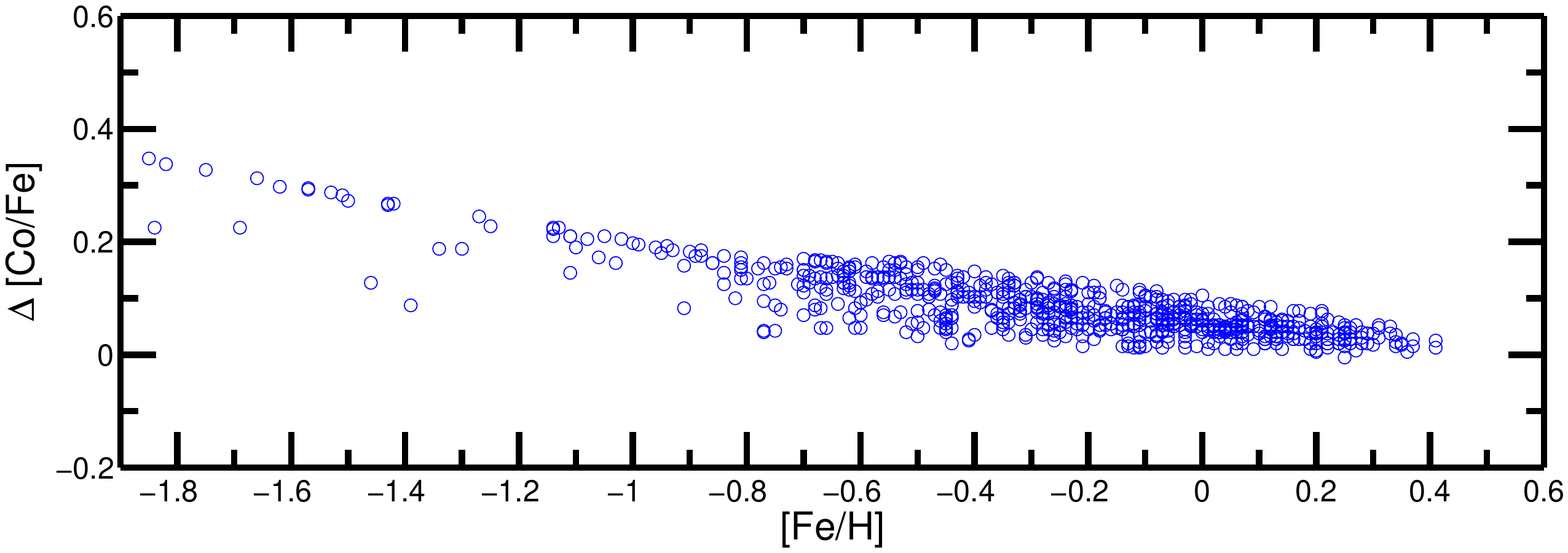}}
\caption{Relation between NLTE corrections for Mn (first column) and Co (second column) with the stellar parameters of our stars, $\teff$, $\log g$, and [Fe/H] (from top to bottom). \label{fig:NLTE_corrections}}
\end{figure*}

\begin{figure*}[ht]
\centering
\resizebox{0.6\hsize}{!}{
\includegraphics[bb=0 60 792 300,clip]{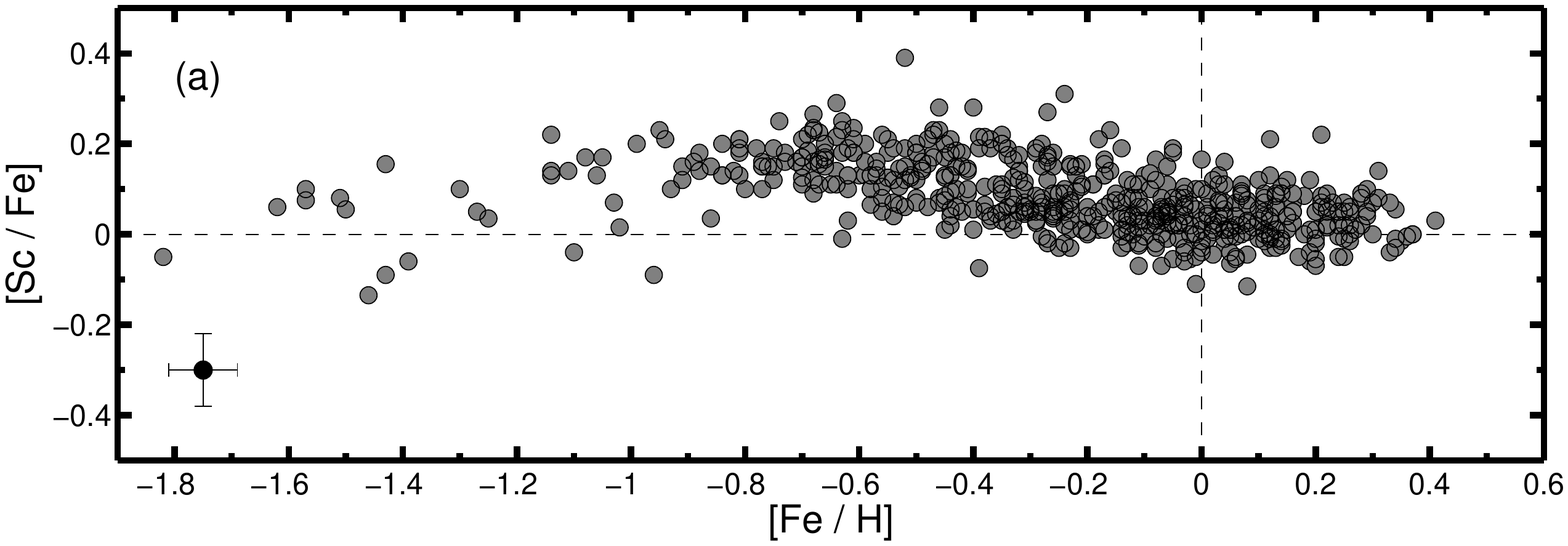}}
\resizebox{0.6\hsize}{!}{
\includegraphics[bb=0 60 792 290,clip]{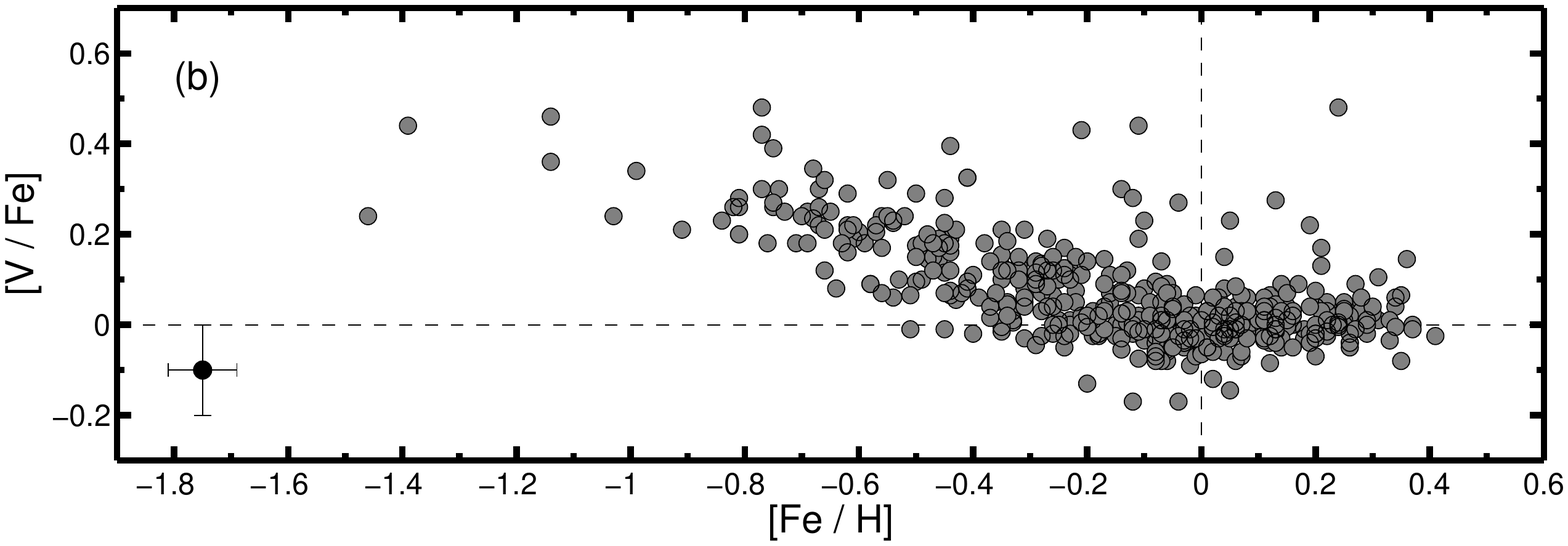}}
\resizebox{0.6\hsize}{!}{
\includegraphics[bb=0 60 792 500,clip]{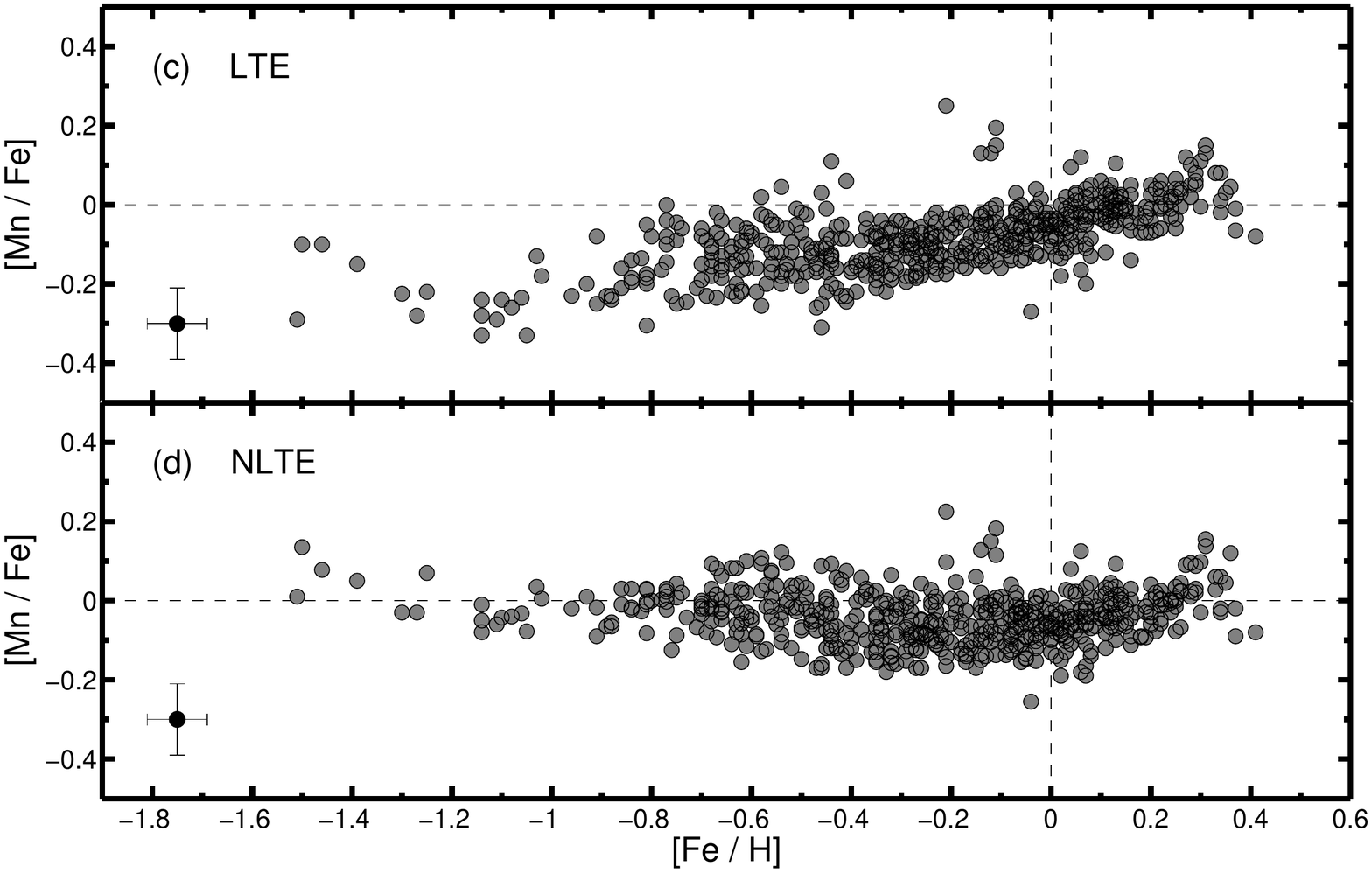}}
\resizebox{0.6\hsize}{!}{
\includegraphics[bb=0 20 792 500,clip]{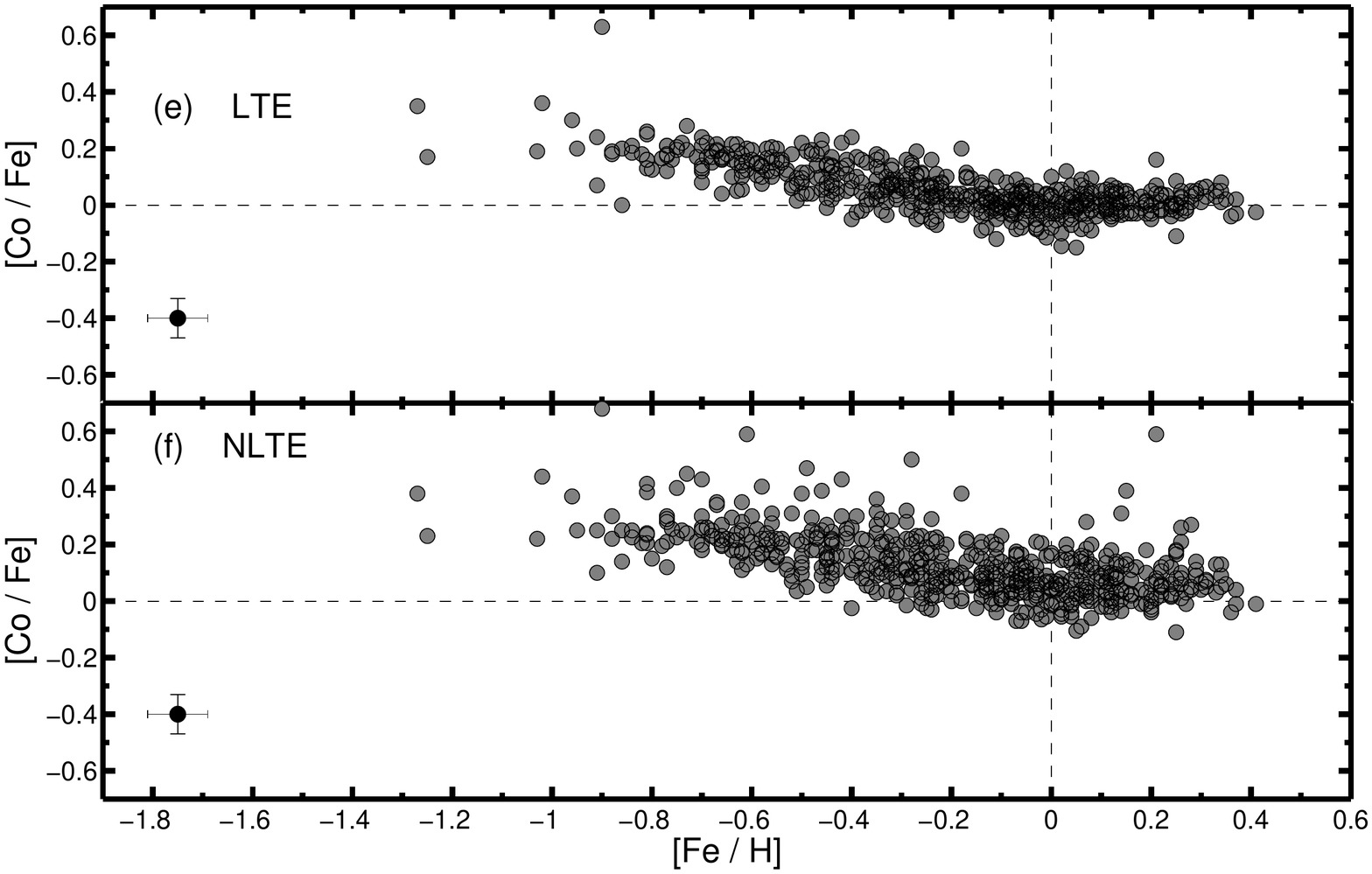}}
\caption{Abundance trends for the complete sample of stars with good fits. (d) and (e) show the Mn results without and with NLTE corrections applied, respectively, as (f) and (g) for Co. In the bottom left  corner the typical error bar is shown. \label{fig:abundance_full}}
\end{figure*}

\begin{figure*}[ht]
\centering
\resizebox{0.6\hsize}{!}{
\includegraphics[bb=0 225 792 450,clip]{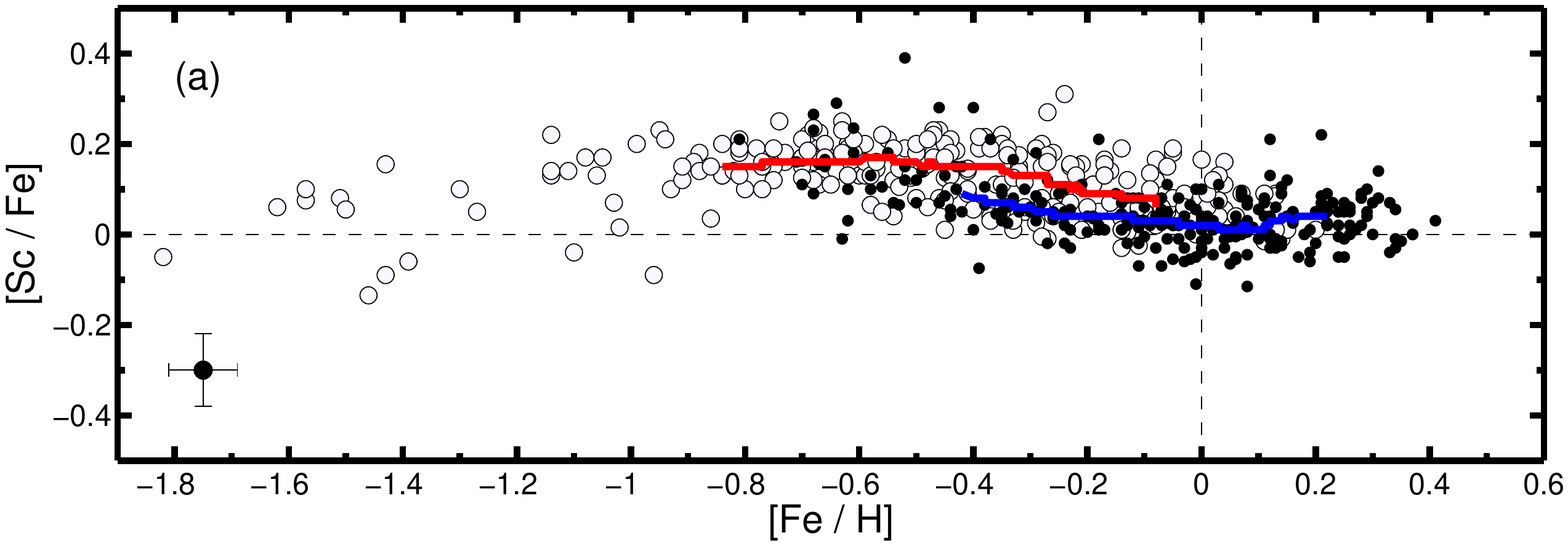}}
\resizebox{0.6\hsize}{!}{
\includegraphics[bb=0 225 792 450,clip]{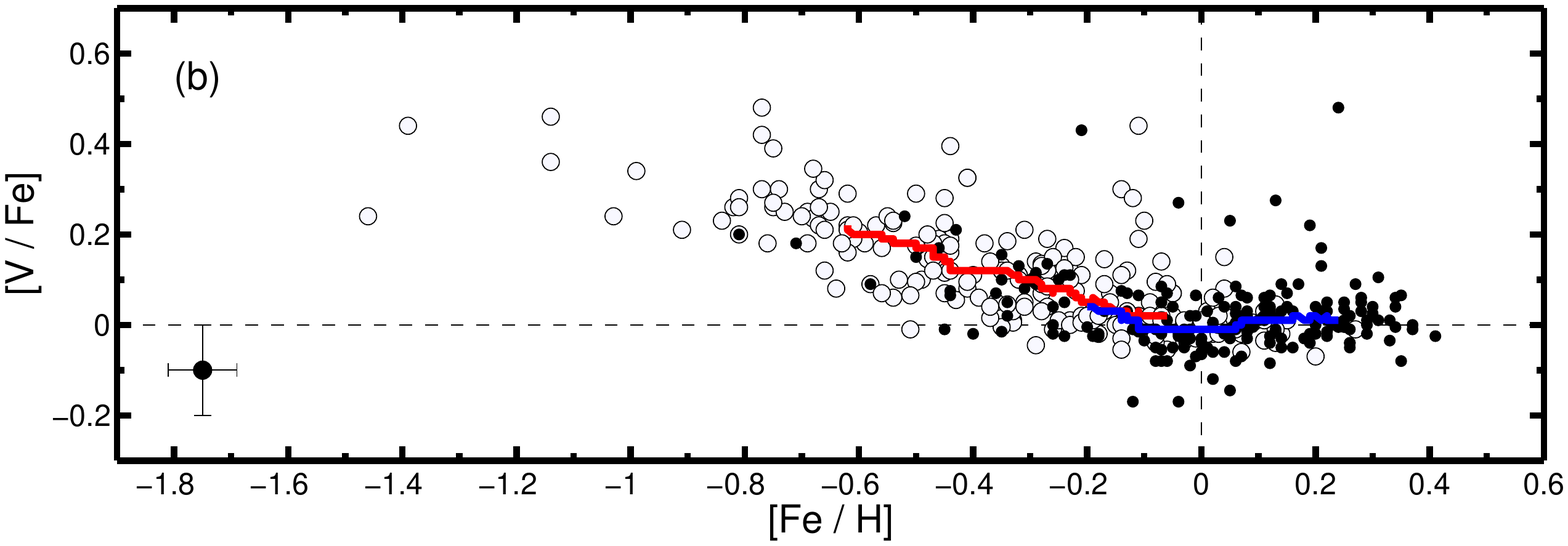} }
\resizebox{0.6\hsize}{!}{
\includegraphics[bb=0 60 792 500,clip]{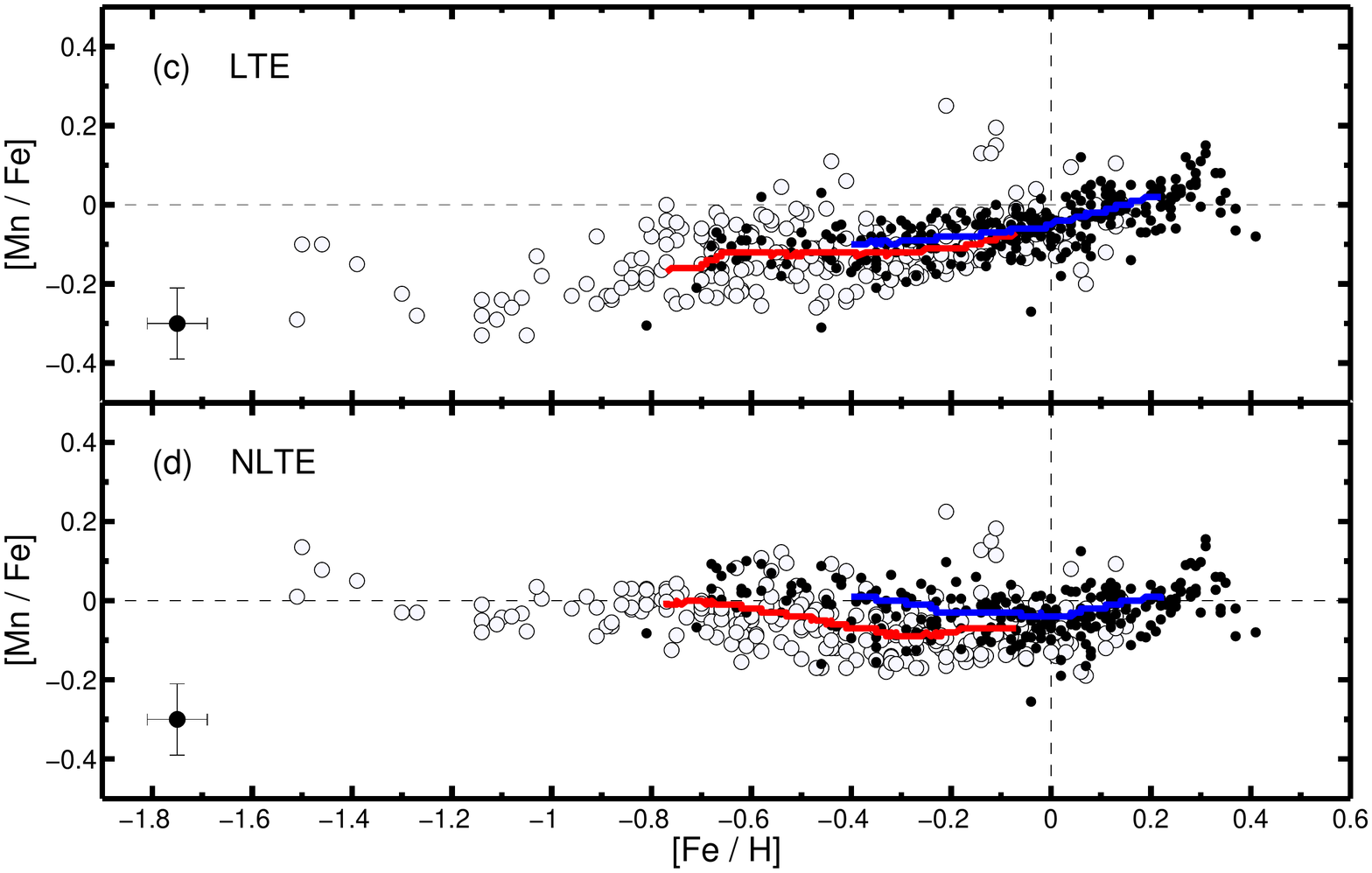}}
\resizebox{0.6\hsize}{!}{
\includegraphics[bb=0 20 792 500,clip]{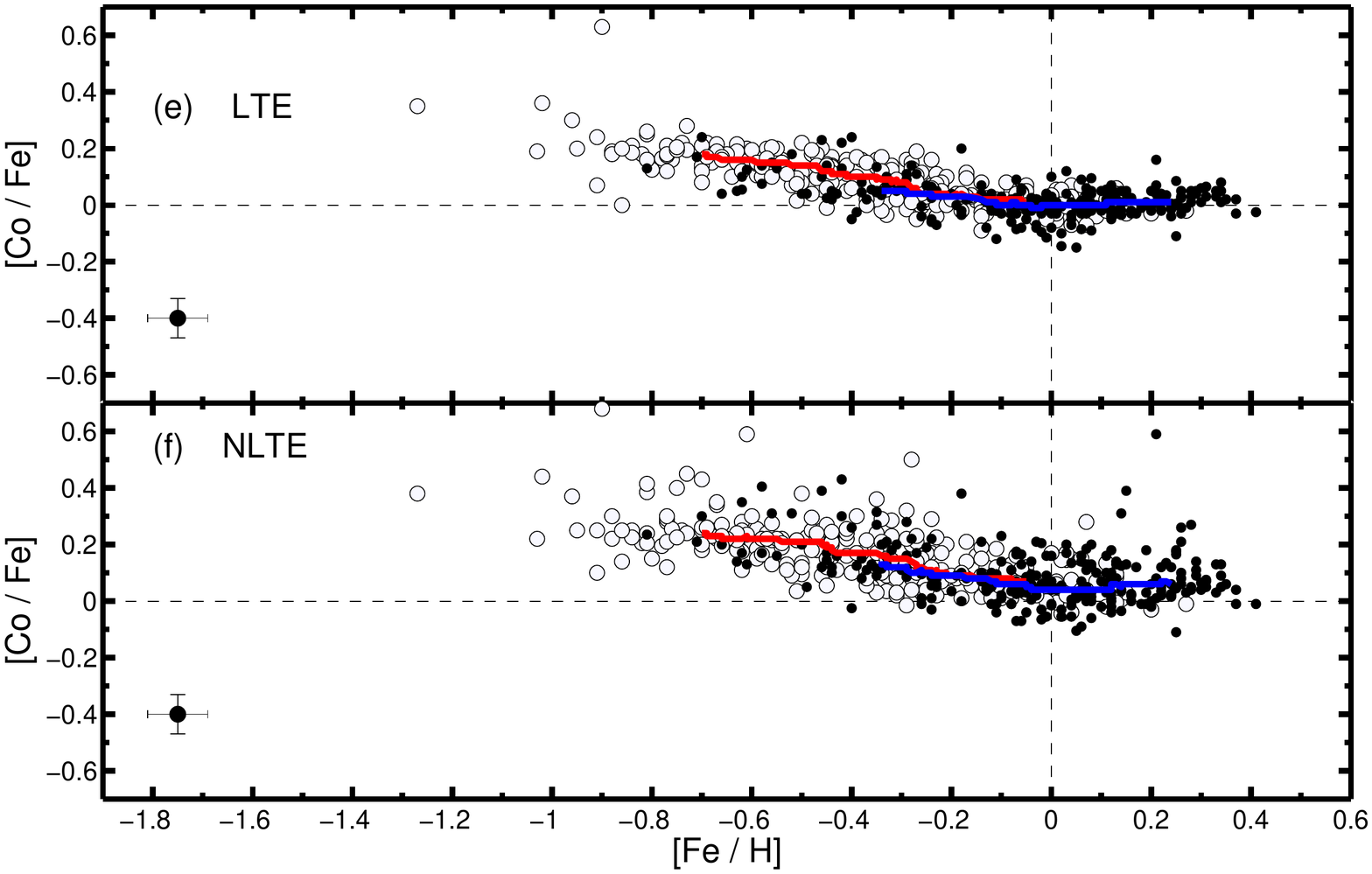}}
\caption{Trends of the abundances of the iron peak elements analyzed compared to the metallicity. Large white dots are stars older than 9\,Gyr, while small black dots are stars younger than 7\,Gyr. (d) and (e) show the Mn results without and with NLTE corrections applied, respectively, as (f) and (g) for Co. The blue and red lines represent the running median of thin and thick disk stars.\label{fig:old_young}}
\end{figure*}

\begin{figure*}[ht]
\centering
\resizebox{0.6\hsize}{!}{
\includegraphics[bb=0 40 792 270,clip]{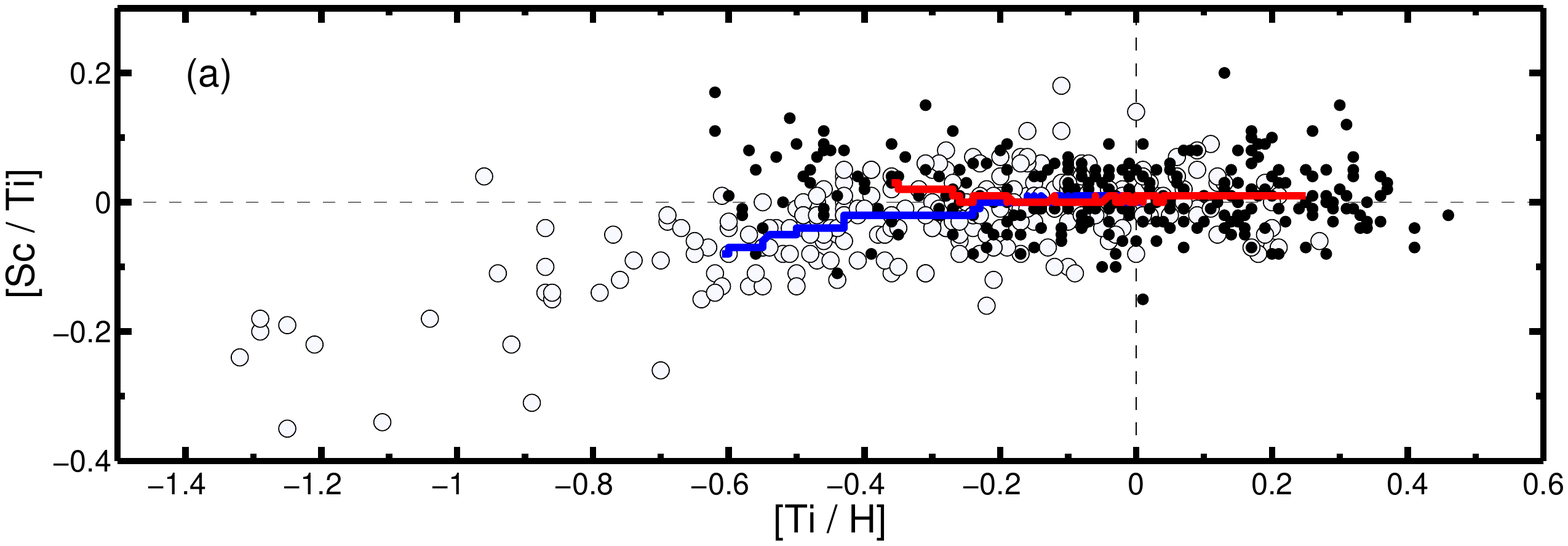}}
\resizebox{0.6\hsize}{!}{
\includegraphics[bb=0 40 792 270,clip]{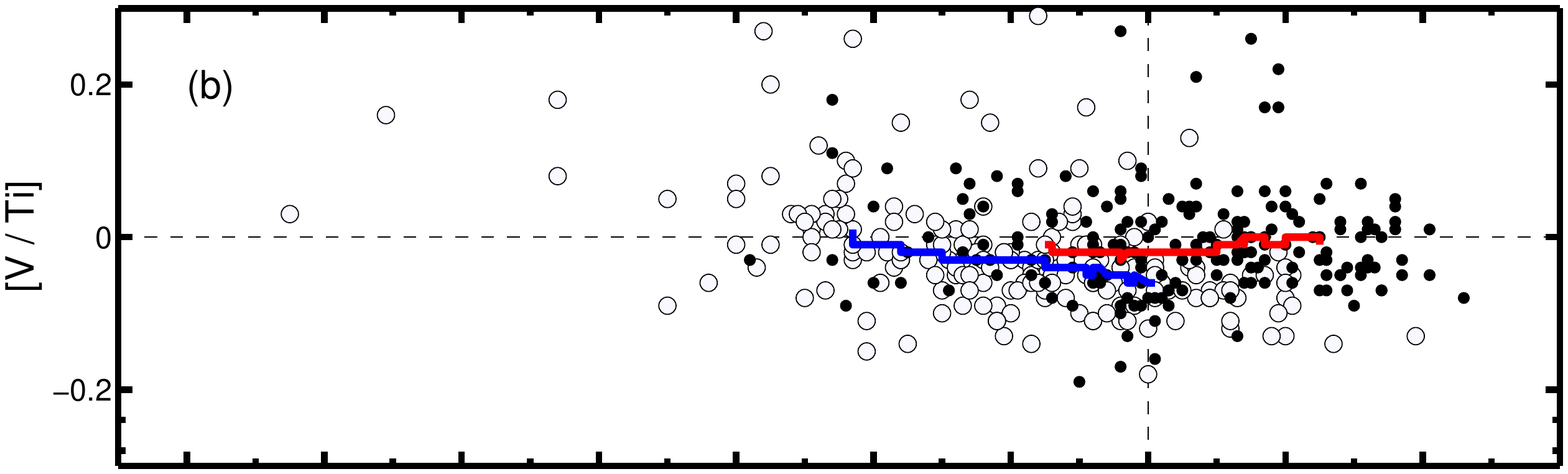} }
\resizebox{0.6\hsize}{!}{
\includegraphics[bb=0 40 792 480,clip]{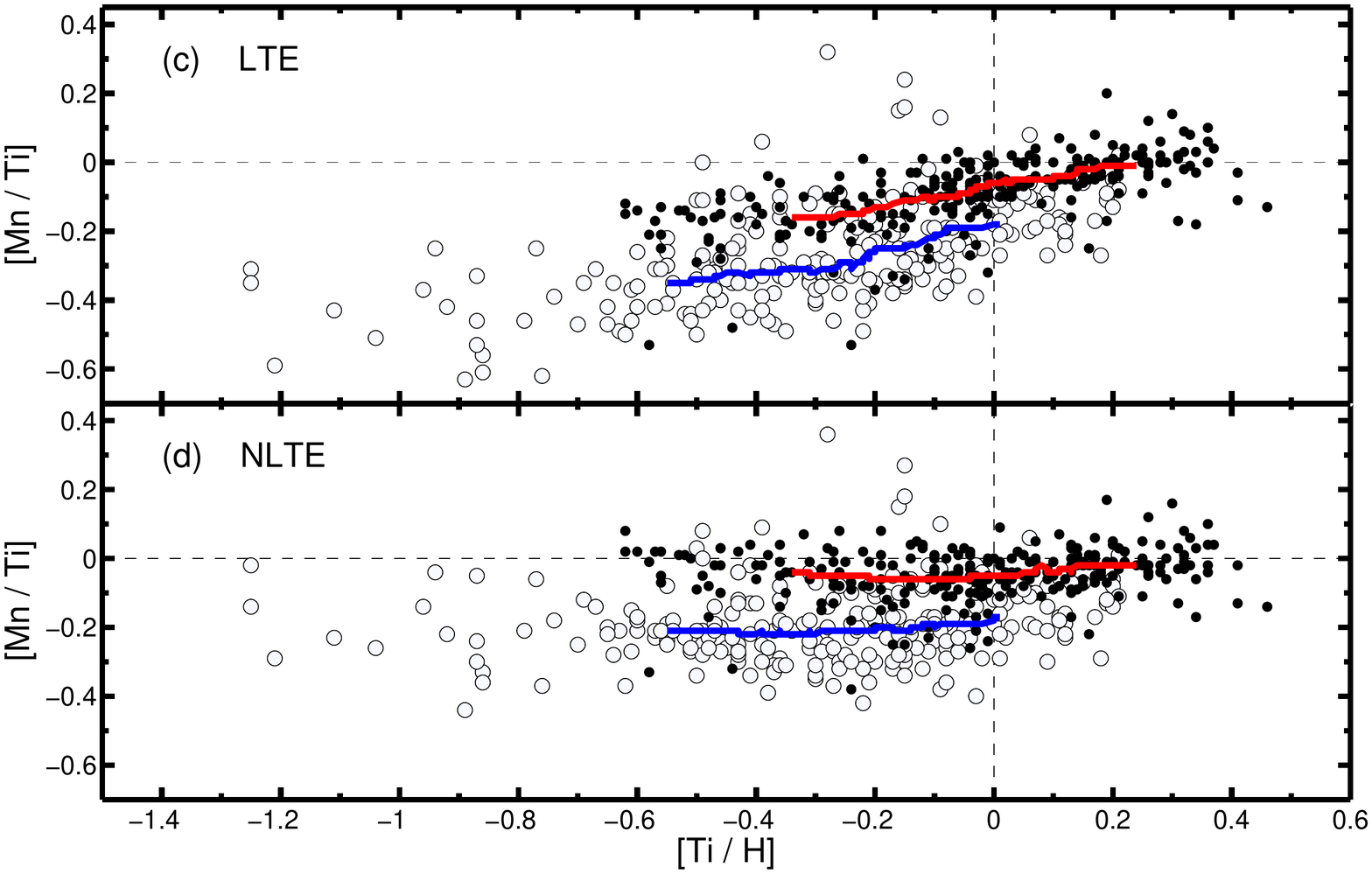}}
\resizebox{0.6\hsize}{!}{
\includegraphics[bb=0 10 792 480,clip]{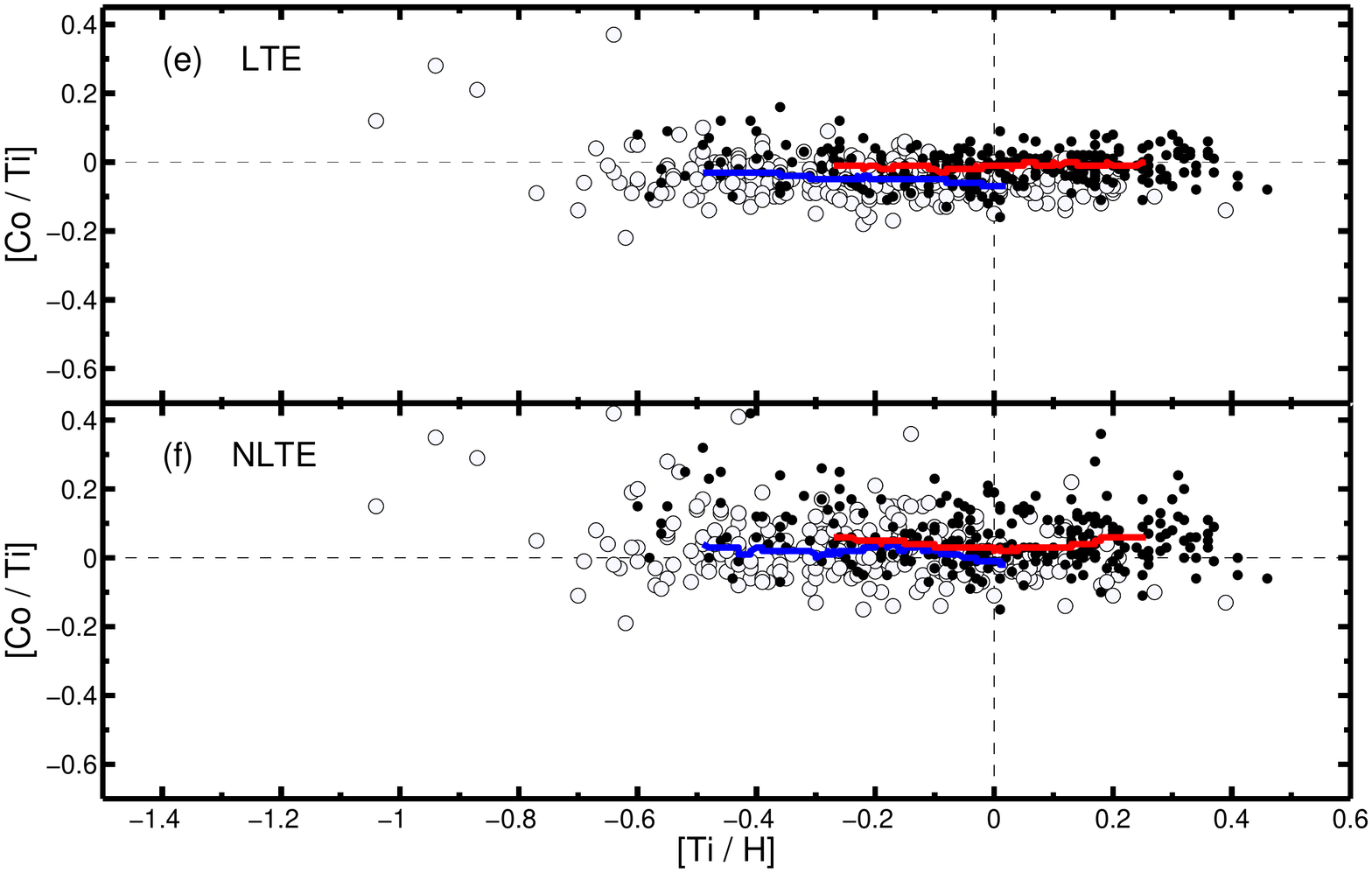}}
\caption{[X/Ti] vs [Ti/H] abundance trends (where X is Sc, V, Co, and Mn respectively). Large white dots are stars with typical kinematics of thick disks, while small black dots are stars with typical kinematic of thin disk. (d) and (e) show the Mn results without and with NLTE corrections applied, respectively, as (f) and (g) for Co. The blue and red lines represent the running median of thin and thick disk stars.\label{fig:all_Ti}}
\end{figure*}

\begin{table}[h!]\tiny
\caption{Uncertainties in stellar parameters and abundance ratios. \label{tab:median_error} \tablefootmark{$^\dagger$}}
\setlength{\tabcolsep}{2 mm}
\centering
\begin{tabular}{c c}
\hline
\\
\multicolumn{2}{c}{Error on parameters} \\
\\
\hline
\\
$\sigma \teff$ [K] & 51\\
$\sigma \log g$ & 0.07\\
$\sigma$ [Fe/H] & 0.06\\
$\sigma \mathrm{\xi_{t}}$ & 0.08\\
$\sigma $[Sc/H]  & 0.08 \\
$\sigma $[V/H]  & 0.10\\
$\sigma $[Mn/H]  & 0.09 \\
$\sigma $ [Co/H]  & 0.07\\
\\
\hline
\end{tabular}
\tablefoot{\tablefoottext{$\dagger$}{Mean standard errors for stellar parameters from \cite{Bensby2014}. Errors for the iron-peak elements are derived from our analysis.}}
\end{table}
\begin{table*}\tiny
\caption{Comparisons of stellar parameters and abundances for stars in common to other studies.\label{tab:error_literature}\tablefootmark{$^\dagger$}}
\setlength{\tabcolsep}{2mm}
\centering
\begin{tabular}{l r r r r}
\hline
\hline
\\
& Reddy03/06 & Adibekyan12 & Takeda07 & Feltzing07\\
\\
\hline
\\
$\Delta \teff$ [K] & +124 $\pm$ 57 (64) & $-$10 $\pm$ 42 (168) & 62 $\pm$ 90 (32) & +54 $\pm$ 71 (95)\\
$\Delta \log g$ & $-$0.05 $\pm$ 0.10 (64) & $-0.06$  $\pm$ 0.10 (168) & 0.05 $\pm$ 0.15 (32) & 0.11 $\pm$ 0.13 (95)\\
$\Delta$[Sc/Fe] [dex] & 0.02 $\pm$ 0.09 (37)& $-0.01$ $\pm$ 0.05 (144) & $-0.03$ $\pm$ 0.05 (20) & -\\
$\Delta$[V/Fe] [dex] & 0.03 $\pm$ 0.12 (28) & $-0.01$ $\pm$ 0.15 (111) & 0.08 $\pm$ 0.23 (20) &-\\
$\Delta$[Mn/Fe] [dex] & 0.09 $\pm$ 0.10 (41) & 0.01 $\pm$ 0.09 (143) & 0.01 $\pm$ 0.09 (14) & $-0.03$ $\pm$ 0.06 (54)\\
$\Delta$[Co/Fe] [dex] & 0.03 $\pm$ 0.12 (36) & $-0.03$ $\pm$ 0.08 (138) & $-0.03$ $\pm$ 0.12 (21) & -\\

\\
\hline
\end{tabular}
\tablefoot{\tablefoottext{$\dagger$}{Differences in stellar parameters from \cite{Bensby2014} and abundances from this work with the stars in common with \cite{Reddy2003,Reddy2006}, \cite{Adibekyan2012}, and \cite{Feltzing2007}. The differences are given as this work minus the other studies. In parentheses, the number of common stars used for the different comparisons. }}
\end{table*}
The analysis was done strictly differentially with respect to the Sun. The solar abundances derived in this way were then subtracted on a line-by-line basis from the abundances of the stars. As shown in \cite{Bedell2014}, different solar spectra observed with different instruments can have differences in derived abundances of up to 0.04\,dex. For this reason, even if the solar values we determine from the different spectra are very similar, we used the proper solar spectrum for each run. As can be seen  in Table 1 in \cite{Bensby2014}, some observational runs do not have their own solar observation. In these cases we used the mean values from all MIKE solar spectra for MIKE run with $R=55\,000$ and the mean value of all the solar spectra for HARPS and FIES observations.

At the end of the fitting of  each spectral line a visual inspection was performed to evaluate if the value for the best abundance from each line did accurately reproduce  the shape of the observed line, as can be seen in Fig.~\ref{fig:good_fit}. Possible causes for a bad fit could be related to the weakness of the line, especially at low metallicities or because of defects in the spectra. The number of stars with poorly fitted lines was higher for V (because of line weakness for metal-poor stars) and was lower for Sc (since lines are usually stronger). Out of a total of 714 stars, we have 594 well fitted stars for Sc, 466 for V, 569 for Mn, and 567 for Co. In Table~\ref{tab:lines} we give the abundances for individual lines derived from the different solar spectra. The same information for the analyzed stars are available in Table~\ref{tab:abundances_single_lines}.

\subsection{Systematic and random error estimation}
To estimate the random errors we performed an analysis to see how the errors in the stellar parameters affect the final abundance for the different elements. Instead of estimating uncertainties for each star, which would have been too time consuming, we decided to select some stars in different parts of the stellar parameter space. The random errors in the stellar parameters calculated in \cite{Bensby2014} using the method from \cite{Epstein2010} were then applied to the  previously selected stars. Then,  the difference with the values obtained without the addition of the errors in the stellar parameters was calculated for each element,  and these differences were considered the errors on the abundance. All the errors for each element were then used to calculate the square mean error. The typical errors for the abundances are around 0.1 dex. The  abundance errors are listed in Table~\ref{tab:median_error}, together with the errors in stellar parameters.

As the analysis is differential relative to the Sun, systematic errors should largely cancel out. To investigate possible systematic offsets we compare our results for stars in common with other studies. Table~\ref{tab:error_literature} shows the differences in stellar parameters and abundances for the stars in common with \cite{Reddy2003,Reddy2006}, \cite{Adibekyan2012}, \cite{Takeda2007}, and \cite{Feltzing2007}. The number of stars in common depends on the element (see Sect.~\ref{sect:abun_det}) while for stellar parameters the stars in common are counted over the entire original sample. \cite{Bensby2014} compared the stellar parameters and abundances for stars in common with other works in literature finding good agreement. In addition, also our abundance results agree with those of \cite{Reddy2003,Reddy2006}, \cite{Adibekyan2012}, \cite{Takeda2007}, and \cite{Feltzing2007}. It is important to notice the difference in stellar parameters for \cite{Feltzing2007} that most likely is due to updated methodologies for determining stellar parameters and abundances in \cite{Bensby2014} compared to \cite{Bensby2005}.

\subsection{NLTE investigation \label{sect:NLTE}}
As mentioned in Sect.~\ref{sect:syn_creation}, SME uses MARCS model atmospheres that are calculated under a standard LTE approximation, where the fundamental assumption is that particle collisions establish the energy distribution of matter. This is a simplification of what is happening inside the star and differences start to be particularly severe close to the stellar surface where the radiation field becomes non-local, anisotropic, and strongly non-Planckian \citep{Bergemann2014}.  

Several studies have investigated NLTE effects for different odd-Z iron-peak elements. 
Scandium was studied in \cite{Zhang2008}, but their study was only done for the Sun. They found that strong NLTE effects are visible only for \ion{Sc}{i} lines while for \ion{Sc}{ii} the effects are negligible.They found that in LTE calculation, log$\varepsilon$ (\ion{Sc}{i}) = 2.90 $\pm$ 0.09 and  log$\varepsilon$ (\ion{Sc}{ii}) = 3.10 $\pm$ 0.05 in the Sun, while for  NLTE calculation their results are log$\varepsilon$ (\ion{Sc}{i}) = 3.08 $\pm$ 0.05 and  log$\varepsilon$ (\ion{Sc}{ii}) = 3.07 $\pm$ 0.04. We could not find any other study of \ion{Sc}{i} and \ion{Sc}{ii} formation in NLTE in the literature and hence will only report LTE results for this element.

Unfortunately we could not find any work on possible NLTE effects for V.

 \cite{Bergemann2008} studied NLTE abundances of Mn in a sample of metal-poor stars, showing that NLTE effects become more pronounced with increasing effective temperature and decreasing metallicity, while changes in gravity start to have effects only at [Fe/H]$\le-2.0$ and $\mathrm{T_{eff}}\ge$ 6000\,K.

Cobalt was studied in \cite{Bergemann2010} who found that the main stellar parameter that determines the sign and magnitude of NLTE corrections is metallicity; for example, for \ion{Co}{i} lines the corrections can vary from +0.1 to +0.6\,dex depending on the metallicity and $\teff$, with higher corrections for more metal-poor and hot stars.

In Fig.~\ref{fig:NLTE_corrections} we show the corrections for Mn and Co for  the whole sample as a function of $\teff$, $\log g$, and [Fe/H]. It is clear that for Mn, $\teff$, and $\log g$ there is no evident trend, which is instead evident for [Fe/H] where it is clear that corrections are higher for metal-poor stars. In the case of Co together with [Fe/H], NLTE corrections are also clearly correlated with $\teff$. 

The question of whether  to use NLTE corrections for Mn and Co, is still open. For Fe-peak elements the main uncertainty is the free parameter $\mathrm{S_{H}}$, the inelastic collision strength between the element under study and neutral hydrogen. One way to calibrate this parameter can be done using the excitation balance of neutral lines of the element being studied from stars with well-known $\teff$. Another way is to use the ionization balance of neutral and ionized lines of the element under investigation from stars with good $\log g$ values. The problem is that calibration using the excitation balance from $\teff$ (as in the case of \citealt{Bergemann2008}) is more uncertain owing to higher uncertainties on $\teff$ for metal-poor stars and the high sensitivity to 3D effects that makes the 1D calibration not necessarily appropriate.

The final decision was to use NLTE corrections for the derived LTE abundances wherever it was possible. We obtained NLTE corrections for our sample (Bergemann, private communication) for Mn and Co as shown in Table~\ref{tab:abundances_NLTE}. We decided to keep the results for both LTE and NLTE since NLTE corrections are rarely used in the literature, especially in the case of Mn.

\section{Results}\label{sect:results}

\subsection{General abundance trends}

\subsubsection{Scandium}\label{sect:Sc}

Figure~\ref{fig:abundance_full}a shows the abundance result for Sc;   it is evident that the [Sc/Fe] - [Fe/H] trend is similar to what is seen for the $\alpha$-elements in \cite{Bensby2014}, i.e., a flat part at solar and supersolar metallicity and then a rise in [Sc/Fe] at $\rm [Fe/H]\approx-0.2$ ending at $\rm [Fe/H]\approx-0.4$ followed by another flat part down to $\rm [Fe/H] \approx-1$ with $\rm [Sc/Fe]\approx 0.2$. There might be a decrease in the [Sc/Fe] ratio for the metal-poor regime that reaches solar [Sc/Fe] for stars around $\rm [Fe/H] \approx-1.6$. 

Our results are consistent with previous works, including \cite{Prochaska2000}, \cite{Nissen2000}, \cite{Allende2004}, \cite{Reddy2006}, \cite{Brewer2006}, \cite{Adibekyan2012}, and \cite{Ishigaki2013}. It is worth noting that the sample studied in \cite{Ishigaki2013} reaches $\rm [Fe/H] \approx -3$ with solar [Sc/Fe], confirming what we see from the few metal-poor stars analyzed in this work.

\subsubsection{Vanadium}\label{sect:V}

Figure~\ref{fig:abundance_full}b shows our results for vanadium. For solar and supersolar metallicity, stars have $\rm [V/Fe] \approx 0$; then from $\rm [Fe/H] \approx -0.1$ to $[Fe/H] \approx -1$ there is a rise with no evident ending plateau and with low scatter, also resembling  in this case an $\alpha$-element trend.

Comparing our results for V with other studies, good agreement is found with \cite{Prochaska2000}, \cite{Reddy2006}, \cite{Brewer2006}, and \cite{Adibekyan2012}. \cite{Ishigaki2013}, however, found nearly flat solar values for stars below [Fe/H]$\approx -$1, while our few metal-poor stars have high [V/Fe]. Only for $\rm [Fe/H] > -1$ did they derive modest enhancement of $\rm [V/Fe] \approx 0.1$, similar to what is seen in this work.

\subsubsection{Manganese}\label{sect:Mn}
The first obvious output from Fig.~\ref{fig:abundance_full}d, already pointed out by \cite{Gratton1989}, is that [Mn/Fe] vs [Fe/H] seems to work opposite to an $\alpha$-element. Manganese shows a constant decrease toward lower metallicities, with a plateau from $\rm [Fe/H]\approx -0.4$ to $\rm [Fe/H] \approx -0.8$.

 More recent works show the same underabundance of Mn compare to Fe, like \cite{Prochaska2000}, \cite{Reddy2006}, \cite{Brewer2006}, \cite{Feltzing2007}, \cite{Adibekyan2012}, and \cite{Ishigaki2013}. Among the different works the agreement  with our results is very good, even though  in \cite{Adibekyan2012}, for example, the plateau that we observe for $\rm [Fe/H] < -0.4$ is not visible. 

However, if the NLTE results are considered, the [Mn/Fe] trend changes drastically resulting in a basically flat trend at solar [Mn/Fe] value. This NLTE trend for Mn would imply a new perspective on  Mn site of production and Galactic chemical evolution.

\subsubsection{Cobalt}\label{sect:Co}

In Fig.~\ref{fig:abundance_full}e it can be seen that the trend for Co in LTE approximation is  a mixture of what was derived for Sc and V. The [Co/Fe] trend is flat for solar and supersolar metallicities and  there is an increase in [Co/Fe] for decreasing metallicity. The rise is less steep than for V and it stops around $\rm [Fe/H] \approx-0.4$ where the trend becomes flat. Because of the scatter and the few stars available, it is difficult to tell from our sample how the Co trend evolves towards lower metallicities. Figure~\ref{fig:abundance_full}f shows Co values with NLTE corrections applied. The scatter increased compared to the LTE case is probably due to the uncertainties resulting from the method used to calculate NLTE corrections (see Sect.~\ref{sect:NLTE}). For Co, \cite{Ishigaki2013} found a flat trend with $\rm [Co/Fe] \pm 0.2$ in the metallicity range $\rm -2.6 \lesssim [Fe/H] \lesssim -1$. General agreement is also found with \cite{Prochaska2000}, \cite{Allende2004}, \cite{Reddy2006}, \cite{Brewer2006}, and \cite{Adibekyan2012}.

\begin{table*}[t]\tiny
\caption{Stellar parameters and abundance results for the entire sample.\label{tab:abundances}\tablefootmark{$^\dagger$}}
\centering
\begin{tabular}{c c c c c c c c c c c}
\hline
\hline
\\
HIP & $\teff$ & $\log g$ & $\mathrm{[Fe/H]}$ & $\mathrm{\xi_{t}}$ & [Sc/H] & [V/H] & [Mn/H] & Mn NLTE & [Co/H] & Co NLTE \\
\\
\hline
\\
80 & 5856 & 4.1 & -0.59 & 1.12 & -0.39 & - & - & 0.14 & -0.49 & \\
\vdots & \vdots & \vdots & \vdots & \vdots & \vdots & \vdots & \vdots & \vdots & \vdots & \vdots\\
\\
\hline
\end{tabular}
\tablefoot{\tablefoottext{$\dagger$}{The table is only available in electronic form at the CDS}\label{tab:abundances_NLTE}}
\end{table*}

\subsection{Thin and thick disk trends}\label{sect:thin_thick}
Generally, it has been found that the thin disk contains relatively young and kinematically cold stars with low $\alpha$-abundance for given metallicity while the thick disk contains old and kinematically hot stars with high $\alpha$-abundances for given metallicity (\citealt{Feltzing2013} and references therein). A common way to define thin and thick disk stellar samples has been to use kinematics (e.g., \citealt{Bensby2003,Bensby2005}). However, such divisions are problematic owing to the overlapping velocity distributions. \cite{Bensby2014} showed that a kinematically cold disk sample contains many old stars with abundance patterns typical of  thick disks, and that a kinematical  hot sample contains many young stars with abundance patterns typical of  thin disks. Instead, stellar ages seem to be a better separator between thin and thick disks (e.g., \citealt{Bensby2014, Haywood2013}). For our age criterion we consider a thick disk to consist of stars older than 9\,Gyr and a thin disk of stars younger than 7\,Gyr.

Figure~\ref{fig:old_young} shows the results when applying an age selection criterion. In order to highlight differences between the young and old populations, a running median was calculated for all the elements, as is shown in Fig.~\ref{fig:old_young}. The running median helps clarify the trends for the thin and thick disks of Sc and for $\rm -0.6 \lesssim[Fe/H]\lesssim -0.1$ the two trends look different. The old population is more similar to an $\alpha$-element while the young population presents a less steep increase in [Sc/Fe], reaching $\rm [Sc/Fe]\approx 0.2$ at $\rm [Fe/H] \approx -0.5$. Since the visual inspection seems to clearly show two different populations, we decided to look at what  the difference between the two populations would be when taking their median values in a fixed metallicity range. We chose the metallicity bin $\rm - 0.4 < [Fe/H] < -0.2$, deriving $\rm [Sc/Fe] = 0.05$ for the young population and $\rm [Sc/Fe] = 0.13$ for the old one. The formal error (i.e.,  1-$\sigma$/$\sqrt{N}$, where N is the number of stars for the population under examination in the metallicity bin) for the young population is 0.01 dex while for the old population it is 0.01 dex. This simple calculation shows that the two populations can  actually be different considering that the spread does not seem to be  too strong; however, a more sophisticated statistical test should be used to confirm it.

Figure~\ref{fig:old_young}b shows the trends for V where the abundance trend for the young population remains flat for the entire metallicity range, with only a few stars at low metallicity enriched in V. The trend for the old component is basically indistinguishable from the young one, except that it extends farther to lower metallicity. 
Other works in the literature looked for differences in thin and thick disk for these two elements. Using V velocity as primary thin/thick disk selection criterion, \cite{Brewer2006} found different trends for Sc, similar to the results shown in Fig.~\ref{fig:old_young}a, while for V no significant differences were found. \cite{Reddy2006} used kinematics to determine thin and thick disk membership probabilities for their stars but did not find any differences between thin and thick disk abundance trends for Sc and V. On the other hand, \cite{Adibekyan2012} used $\alpha$-enhancement to define thin and thick disk samples, and found that the abundance trends differed for both Sc and V. 

A similar investigation is also presented  for Mn and Co, considering both LTE and NLTE results separately. In Fig.~\ref{fig:old_young}c-d, the  LTE case shows that  the trends for the young and old populations are similar and difficult to distinguish, as also pointed out in \cite{Brewer2006}. In the NLTE case, a difference between old and young population seems to be present in the metallicity range $\rm -0.6 \lesssim [Fe/H] \lesssim -0.1$. We decided to derive the median values for the old and young populations between $\rm -0.4 < [Fe/H] < -0.2$ as was performed for Sc. For the young population the median [Mn/Fe] is $-0.01$\,dex, while for the old population $\rm [Mn/Fe] = -0.09$. The formal error in this case is 0.01 for both the young and old population. As in the Sc case, this simple calculation shows that the two populations do not overlap, but unlike Sc, NLTE Mn appears to be affected by much stronger scatter that makes it difficult to separate the two populations. We note that we do not see the difference in the Mn trends between the thin and thick disk populations reported by \cite{Feltzing2007}. 

Figure~\ref{fig:old_young}e-f show the trends for Co where the running median indicates that there are no clear differences between the two populations.  The NLTE case for Co also does not show any distinction between thin and thick disk.   \cite{Brewer2006} found no distinction, while \cite{Reddy2006} and \cite{Adibekyan2012} found differences between the two populations.

\begin{figure*}[ht]
\centering
\resizebox{0.7\hsize}{!}{
\includegraphics[bb=0 270 750 500,clip]{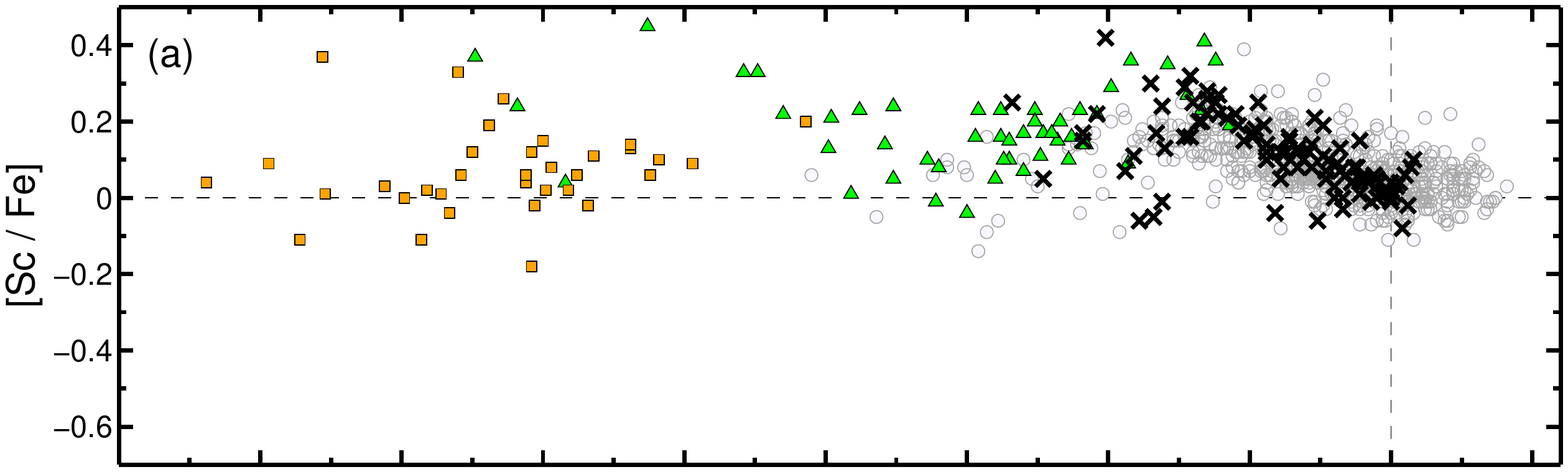}}
\resizebox{0.7\hsize}{!}{
\includegraphics[bb=0 270 750 500,clip]{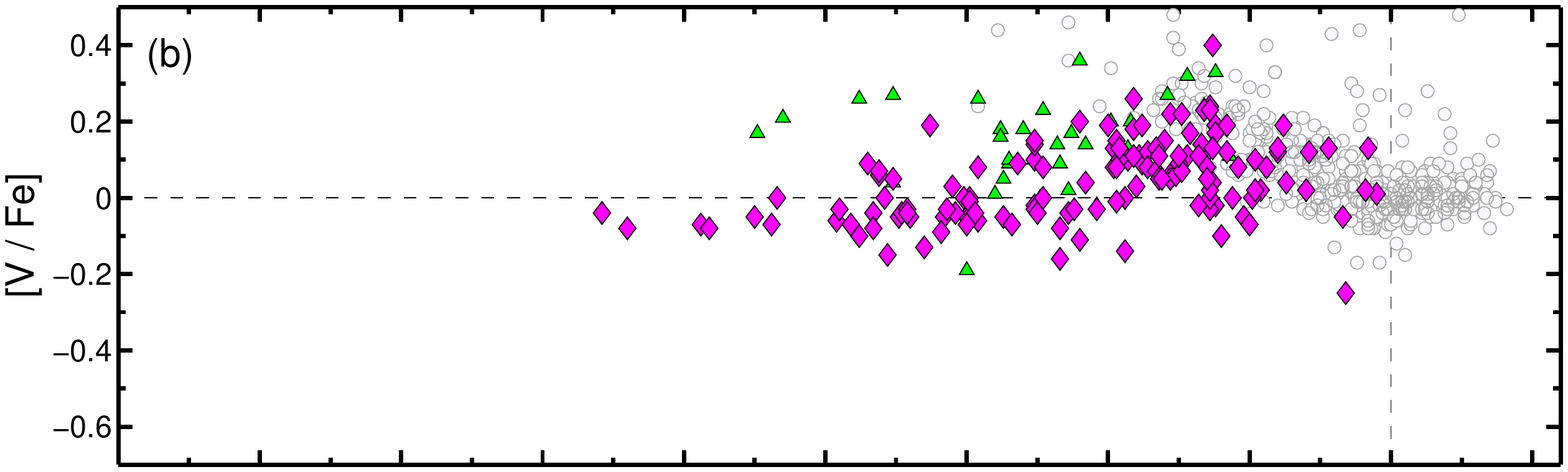}}
\resizebox{0.7\hsize}{!}{
\includegraphics[bb=0 270 750 500,clip]{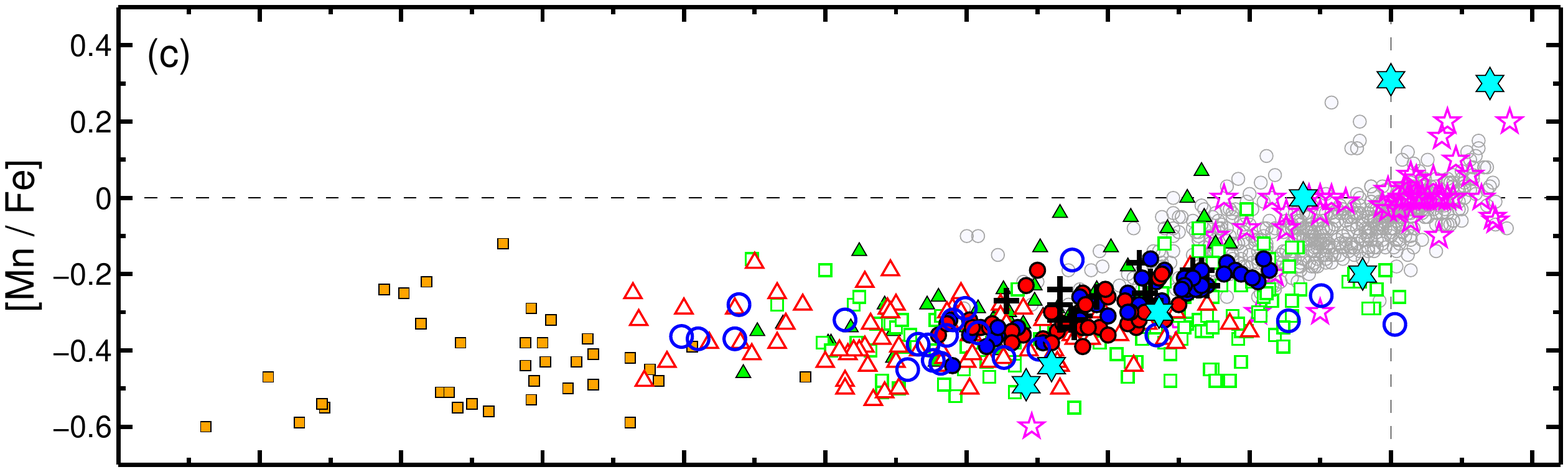}}
\resizebox{0.7\hsize}{!}{
\includegraphics[bb=0 230 750 500,clip]{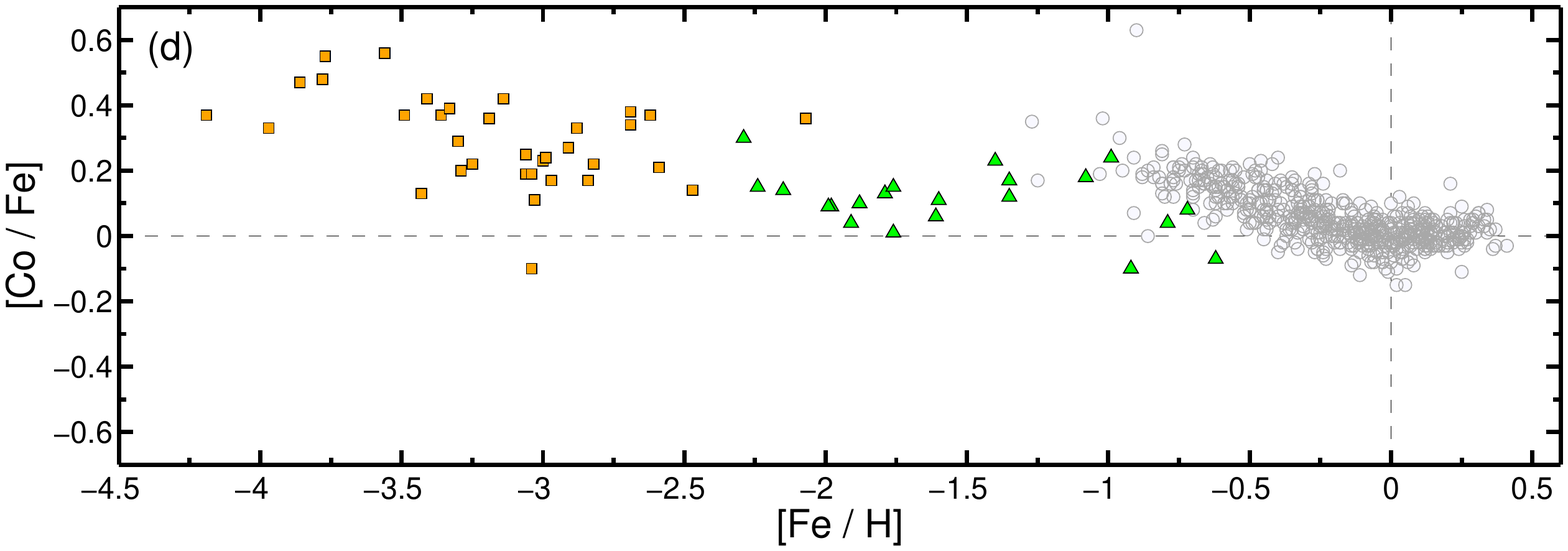}}
\caption{Galactic trends with data from the literature with gray dots showing our results for the Galactic disk. In panel (a) for Sc with data from \cite{Ishigaki2013}  (green filled triangles) and \cite{Nissen2000} (black crosses) dwarfs from thick disk and halo, and metal-poor giants from \cite{Cayrel2004} (orange squares). In (b) V green filled triangles as in panel (a) and magenta filled diamonds from \cite{Fulbright2000} from galactic halo. Panel (c) shows the trends for Mn, with orange squares and green filled triangles as before. In addition, \cite{Nissen2011} thick disk stars (black crosses), high $\alpha$ population (blue filled dots), and low $\alpha$ population (red filled dots), \cite{Barbuy2013} (magenta stars), and \cite{McWilliam2003} (cyan six-pointed stars) from the bulge and \cite{Sobeck2006} from globular clusters (blue circles) and \cite{Sobeck2006} from field stars using stars from \cite{Fulbright2000} (green squares) and \cite{Simmerer2004} (red triangles). In (d) Co trend for \cite{Ishigaki2013} (green filled triangles) and \cite{Cayrel2004} (orange squares).\label{fig:all_literature_galactic}}
\end{figure*}

\subsection{Comparison with $\alpha$-elements}

Next we  explore possible relations between odd-Z iron-peak elements and the $\alpha$-elements Mg, Si, Ca, and Ti. Since the sites of production of $\alpha$-elements are better known, this comparison will shed more light on production sites for odd-Z iron-peak elements. We performed these comparisons with all the $\alpha$-elements studied in \cite{Bensby2014}, but  here we only show the comparison with Ti because there is a  much clearer and more well-defined division between the thin and thick disks in Ti (see \citealt{Bensby2014}).

Titanium is mainly produced in intermediate mass SNII and, in smaller quantities, by SNIa \citep{Tsujimoto1995,Thielemann2002}. 
Figure~\ref{fig:all_Ti} shows the abundance trends of Sc, V, Mn, and Co with Ti as the reference element using the age selection criteria from Sect.~\ref{sect:thin_thick} to highlight possible differences between thin and thick disk trends. As is clear from Fig.~\ref{fig:all_Ti}, the two populations are completely mixed for Sc, V, and Co (LTE and NLTE case). Figure~\ref{fig:all_Ti}b and e-f show the trends for V and Co LTE and Co NLTE, respectively, where it is easy to see that even if the spread is large (especially for V) the two trends are basically flat. This could mean that V and Co, are produced in the same sites as Ti. In Fig.~\ref{fig:all_Ti}a, Sc shows a flat part from  supersolar [Ti/H] down to $\rm [Ti/H] \approx -0.5$ and then a decline in [Sc/Ti] for lower [Ti/H]. This decline is clearly related to the decline to solar Sc value for low metallicity, as seen in Fig.~\ref{fig:abundance_full}a.

The same investigation with Mn shows some interesting trends, since  in the cases of LTE values (Fig.~\ref{fig:all_Ti}d) and NLTE corrections (Fig.~\ref{fig:all_Ti}e), the thin and thick disk results are distinct. Also here the inclusion of NLTE corrections creates a flatter trend with [Ti/H] compared to the LTE values. In the NLTE case this trend can mean that the production sites of Mn are the same as Fe (SNII and SNIa), but with no or weak metallicity dependence. In LTE the rising trend can indicate metallicity dependent Mn yields in SNII \citep{McWilliam2003}.  It is worth noting that in both cases the young population presents higher [Mn/Ti] values, meaning that Mn is produced more than  Ti in thin disk stars. This is not seen in the other elements; it is a specific characteristic of Mn.

\section{The origin of Mn}\label{sect:origin_Mn}
Manganese is often studied because of its peculiar behavior as an Fe-peak element, as  is visible in Fig.~\ref{fig:old_young}d. 
Manganese is thought to be mostly produced by explosive silicon burning in massive stars in their outer incomplete Si-burning layers \citep{Umeda2005} and in SNIa \citep{Iwamoto1999}. If Mn were  over-produced in SNIa compared to Fe, it would be expected that the low $\alpha$ population in \cite{Nissen2011} should be enriched in Mn compared to the high $\alpha$ (see Fig.~\ref{fig:all_literature_galactic}c).  The thin and thick disk [Mn/Fe] trends would also be different owing to the different evolution histories. This is not seen, however, except  in the case when we apply NLTE corrections to the Mn abundances (Fig.~\ref{fig:old_young}). As shown in \cite{Kobayashi2011} and \cite{Nomoto2013} models can reproduce quite well the Mn values at solar and low metallicities. However, there are only a few disk observations in the range $\rm -2.0 < [Fe/H] < -0.8$ (in Fig.~\ref{fig:all_literature_galactic}c most of the observations are for halo plus some data for the bulge), so our results are important to fill this gap, and give useful information when comparing models with observations. Unfortunately, the models cannot reproduce the observed Mn trends when NLTE corrections are taken into account. The basically flat trend observed in the NLTE case would be impossible to explain with current nucleosynthesis models because SNII yields would not permit the trend [? the ratio?] to reach $\rm [Mn/Fe] \approx 0$. In addition, at $\rm [Fe/H] \approx -1$  when SNIa enrichment is important owing to high production of Mn in SNIa, [Mn/Fe] should rise.

In metallicities below $\rm [Fe/H] \approx -1$, the low [Mn/Fe] ratios are mainly determined by the SNII yields from massive stars \citep{Tsujimoto1998}. For $\rm [Fe/H] \gtrsim -1$, the increasing [Mn/Fe]  values with increasing [Fe/H] is interpreted as a contribution from SNIa \citep{Kobayashi2006}. It was also suggested that the dependence of Mn yields from SNIa with metallicity may contribute to the increase in the [Mn/Fe] ratios \citep{Cescutti2008}. The LTE results derived in this study confirm what is expected from the models. Our results also show that Mn trends change quite drastically if NLTE corrections are used, and in this case [Mn/Fe] is basically flat with subsolar values. This means that Mn shares the same production sites as Fe and does not show strong metallicity dependence. We also exclude metallicity dependence for the rise in the LTE case, considering that the metallicity for which [Mn/Fe] increases is basically the same for which [Sc, Co/Fe] start to decline, i.e., the prompt of SNIa explosions.

\section{The origin of Sc, V, and Co}\label{sect:origin_other}
In the theoretical calculation of \cite{Woosley1995}, both Sc and V are produced through explosive oxygen, silicon, and neon burning in massive stars 
The yields for Sc in the \cite{Woosley1995} model show a tendency to increase with increasing metallicity. As  seen  in Fig.~\ref{fig:abundance_full}a, the [Sc/Fe] ratio decreases to solar values at $\rm [Fe/H]\approx -1$. This is also visible in Fig.~\ref{fig:all_literature_galactic}a from other works in other stellar populations. This decrease is also present  in theoretical models owing to metallicity dependencies of Sc yields at $Z > 0.001$ for Type II supernovae/hypernovae \citep{Nomoto2013}. However, these models have been unable so far  to reproduce the observed abundance of Sc \citep{Nomoto2013}. Models predict under-abundant [Sc/Fe], but this difference between observations and simulation could be caused by the strong dependence that yields have with the parameters of supernova explosions. Our results for Sc confirm its behavior  as an $\alpha$-element, produced probably in the same sites as Ti. The decline in [Sc/Fe] at $\rm [Fe/H] \approx -0.5$ could then be due to progressively higher rate of SNIa enrichment.

Similar results are seen for V (for V abundances in other galactic populations, see Fig.~\ref{fig:all_literature_galactic}b), which in models from \cite{Nomoto2013} [V/Fe] is under-abundant compared to observations. As in the case of Sc, the parameters of the explosions play a role and small differences in these parameters could create different yields meaning big differences when comparing observations with predictions. 
Also in this case the tuning of the explosion could play a role and this could explain the big difference between real observations and theoretical results. In addition to this it seems that V yields are only slightly dependent on metallicity, because of the smoother rise in [V/Fe]. 

Cobalt is produced in both SNII and SNIa \citep{Timmes1995,Kobayashi2006}. A chemical evolution model of \cite{Timmes1995} suggests that the trend in [Co/Fe] with respect to [Fe/H] can be explained by the dependence of Co ejection on mass and metallicity of massive stars and by the production of Fe through SNIa. At low metallicities, hypernovae (stars with $\gtrsim$\,20\,M$_{\odot}$ but that explode with energy about ten times higher than normal SNII) could have played an important role in the chemical enrichment of the ISM because of a higher star formation rate, implying that more massive stars were present \citep{Nomoto2006}. \cite{Nomoto2013} show that the higher energy for explosions of hypernovae tend to produce larger [Co, V/Fe] and smaller [Mn/Fe]. This model could explain our trends in metal-poor regime for Co, but also for V and Mn together with the high [Co/Fe] that is visible in very metal poor regimes, as shown in Fig.~\ref{fig:all_literature_galactic}d.

\section{Summary and outlook}\label{sect:summary}
We have determined, through spectral synthesis, abundances for the odd-Z elements Co, V, Mn, and Sc for a large sample of solar neighborhood F and G dwarf stars in order to investigate their formation sites and their evolution in the Galactic disk. In total we determined Sc abundance for 594 stars, V abundance for 466 stars, Mn abundance for 569 stars, and Co abundance for 567 stars. In summary, our main findings and conclusions are as follows:
\begin{itemize}
\item Scandium is categorized as an iron-peak element, but its abundance characteristic places it as an element like Mg. This is visible in the trend when compared to Fe and also when compared with Ti, showing that it shares the same production sites. Theoretical models predict this together with a metallicity dependence of the yields, but the exact amount of produced Sc is very uncertain because yields are strongly dependent on SNII explosion parameters.
\item Vanadium is thought to be produced in the same sites as Sc and this is confirmed by our observations, even though we cannot reach metallicity below $\rm [Fe/H] \approx -1$. A comparison with Ti shows that they might have the same production sites. As in the Sc case, models cannot reproduce the observations, probably owing to strong dependence on explosions parameters.
\item Manganese presents a completely different behavior compared to the other odd-Z iron-peak elements, showing subsolar values for all metallicities up to solar and supersolar [Fe/H], where it presents solar [Mn/Fe]. This trend can be explained by models where Mn is produced in SNII at low metallicity while at $\rm [Fe/H] \approx -1$ it is produced by SNIa. Applying NLTE corrections to Mn corrections, [Mn/Fe] becomes basically flat over the metallicity range that our stars span, Mn shares the same production sites as Fe and is produced in the same quantity. If NLTE corrections are taken into account, no models have been able to  explain this behavior.
\item Cobalt shows a trend similar to $\alpha$-elements and it seems that [Co/Fe] is also over-abundant  at low metallicities. Chemical models  cannot explain this trend, unless Co could be produced by hypernovae at low metallicity. In addition, it is possible that increasing inhomogeneous enrichment due to single SN at very low metallicities could explain this high [Co/Fe].
\item We also investigated potential differences between thin and thick disks. We did not find any evident differences in abundance trends, apart from small deviations in metallicity range $\rm -0.6 \lesssim [Fe/H] \lesssim -0.2$ for Sc and Mn when NLTE corrections are applied.
\end{itemize}

Using the same star sample and the same methodology we are analysing some r- and s-process elements, like Sr, Zr, La, Ce, Nd, Sm, and Eu. In addition, we will derive abundances for the same iron-peak and r- and s- process elements in a sample of microlensed dwarf stars in the Galactic bulge \citep{Bensby2013}.

\subsection*{Acknowledgements}
T.B. was funded by grant No. 621-2009-3911 from the Swedish Research Council. We thank Maria Bergemann for calculating NLTE corrections for our stars. We would also like to thank the anonymous referee for useful comments that helped to improve the paper.

\bibliographystyle{aa} 
 \bibliography{Chiara_references} 
\newpage
\begin{appendix}

\section{Line lists}

\begin{table*}[ht]\tiny
\caption{Line lists for the synthesis of \ion{Sc}{ii}.\label{tab:hfs1}\tablefootmark{$^\dagger$}}
\begin{center}
\setlength{\tabcolsep}{1 mm}
\begin{tabular}{c c c | c c c | c c c | c c c | c c c }
\hline
\multicolumn{3}{c|} {\ion{Sc}{ii}  5526  } & \multicolumn{3}{c|}{\ion{Sc}{ii} 5657 } & \multicolumn{3}{c|}{ \ion{Sc}{ii} 5657} & \multicolumn{3}{c|}{ \ion{Sc}{ii} 5684} & \multicolumn{3}{c}{ \ion{Sc}{ii} 6245} \\
Element & $\lambda$ ({\AA}) & log(gf)  & Element & $\lambda$ ({\AA}) & log(gf) & Element & $\lambda$ ({\AA}) & log(gf) & Element & $\lambda$ ({\AA}) & log(gf) & Element & $\lambda$ ({\AA}) & log(gf)\\
\hline
\ion{Ca}{i} & 5525.695 & -1.153 & \ion{V}{i} & 5656.881 & -1.011 & \ion{Si}{i} &  5666.677 & -1.797 & \ion{Na}{i} & 5682.633 & -0.706 & \ion{Si}{i} & 6244.466 & -1.091\\
\ion{Fe}{i} & 5525.848 & -1.574 & \ion{V}{i} & 5657.435 & -1.020 & \ion{Fe}{i} & 5666.788 & -3.385 & \ion{Fe}{i} & 5683.452 & -3.406 & \ion{Ni}{i}  & 6244.965 & -1.265\\
\ion{Fe}{i} & 5526.203 &  -2.353 & \ion{Fe}{i} & 5657.665 &  -2.699 & \ion{Ni}{i}  & 5666.793 & -2.262 & \ion{Ce}{ii} & 5683.745 & -0.860 & \ion{V}{i} & 6245.219 & -2.005\\
\ion{Sc}{ii} & 5526.770 & -2.665  & \ion{Sc}{ii} & 5657.886 & -1.126 & \ion{Sc}{ii} & 5667.136 & -1.834 & \ion{Sc}{ii} & 5684.190 & -1.603 & \ion{Sc}{ii} & 6245.621 & -1.736  \\
\ion{Sc}{ii} & 5526.775 & -2.262  & \ion{Sc}{ii} & 5657.888 & -1.696 & \ion{Sc}{ii} & 5667.141 & -2.030 & \ion{Sc}{ii} & 5684.191 & -2.092 & \ion{Sc}{ii} & 6245.629 & -2.476  \\
\ion{Sc}{ii} & 5526.779 & -2.050  & \ion{Sc}{ii} & 5657.893 & -1.696 & \ion{Sc}{ii} & 5667.148 & -2.030 & \ion{Sc}{ii} & 5684.193 & -2.791 & \ion{Sc}{ii} & 6245.631 & -1.907  \\
\ion{Sc}{ii} & 5526.779 & -1.452  & \ion{Sc}{ii} & 5657.894 & -1.524 & \ion{Sc}{ii} & 5667.154 & -3.215 & \ion{Sc}{ii} & 5684.200 & -1.074 & \ion{Sc}{ii} & 6245.636 & -3.476  \\
\ion{Sc}{ii} & 5526.783 & -1.940  & \ion{Sc}{ii} & 5657.895 & -1.538 & \ion{Sc}{ii} & 5667.157 & -2.034 & \ion{Sc}{ii} & 5684.204 & -1.896 & \ion{Sc}{ii} & 6245.638 & -2.293  \\
\ion{Sc}{ii} & 5526.783 & -1.247  & \ion{Sc}{ii} & 5657.899 & -1.538 & \ion{Sc}{ii} & 5667.163 & -2.034 & \ion{Sc}{ii} & 5684.205 & -1.976 & \ion{Sc}{ii} & 6245.640 & -2.114  \\
\ion{Sc}{ii} & 5526.786 & -1.181  & \ion{Sc}{ii} & 5657.901 & -2.220 & \ion{Sc}{ii} & 5667.167 & -2.289 & \ion{Sc}{ii} & 5684.206 & -2.351 & \ion{Sc}{ii} & 6245.644 & -3.058  \\
\ion{Sc}{ii} & 5526.787 & -1.919  & \ion{Sc}{ii} & 5657.902 & -1.549 & \ion{Ti}{i} & 5667.594 & -0.596, & \ion{Sc}{ii} &                        5684.215 & -2.351 & \ion{Sc}{ii} & 6245.646 & -2.260  \\
\ion{Sc}{ii} & 5526.788 & -1.185  & \ion{Sc}{ii} & 5657.904 & -1.549 & \ion{Fe}{i} & 5667.665 & -3.030, & \ion{Sc}{ii} &                        5684.217 & -2.080 & \ion{Sc}{ii} & 6245.647 & -2.385  \\
\ion{Sc}{ii} & 5526.789 & -0.523  & \ion{Sc}{ii} & 5657.906 & -1.722 & 
\ion{Cr}{i} & 5667.805 & -1.231 & \ion{Si}{i} & 5684.484 & -1.650 & \ion{Sc}{ii} &                                                       6245.650 & -2.824  \\
\ion{Sc}{ii} & 5526.790 & -2.028  & \ion{Sc}{ii} & 5657.906 & -3.646 & & & & \ion{Fe}{i} & 5684.485 & -2.389 &\ion{Sc}{ii} &                                                       6245.651 & -2.319  \\
\ion{Sc}{ii} & 5526.790 & -1.246  & \ion{Sc}{ii} & 5657.908 & -1.722 & & & & \ion{Ce}{ii}  & 5685.839 & -0.170 &\ion{Sc}{ii} &                                                 6245.652 & -2.803  \\
\ion{Sc}{ii} & 5526.791 & -0.644  & \ion{Sc}{ii} & 5657.909 & -1.898 & & & & & & & \ion{Sc}{ii} &                                                      6245.655 & -2.678  \\
\ion{Sc}{ii} & 5526.792 & -1.368  & \ion{Fe}{ii} & 5657.924 & -4.112 & & & & & & & \ion{Sc}{ii} &                                                                                      6245.656 & -2.502  \\
\ion{Sc}{ii} & 5526.793 & -1.551  & \ion{S}{i} & 5657.958 & -1.366 & & & & & & & \ion{Sc}{ii} &                                                                                6245.657 & -2.581  \\
\ion{Sc}{ii} & 5526.793 & -0.780  & \ion{Sc}{ii} & 5658.361 & -1.208 & & & & & & & \ion{Fe}{i} & 6246.318 & -0.733\\
\ion{Sc}{ii} & 5526.794 & -0.936  & & & & & & & & & & \ion{Co}{i}  & 6247.284 & -0.639 \\
\ion{Sc}{ii} & 5526.795 & -1.697  & & & & & & & & & & \ion{Fe}{ii} & 6247.354 & -2.172 \\
\ion{Sc}{ii} & 5526.795 & -1.355  & & & & & & & & & & \\  
\ion{Sc}{ii} & 5526.795 & -1.121  & & & & & & & & & & \\
\ion{Ce}{ii} & 5526.861 & -1.930 & & & & & & & & & & \\
\ion{Fe}{ii} & 5527.354 & -8.134 & & & & & & & & & & \\
\ion{Y}{i} & 5527.548 & 0.471 & & & & & & & & & & \\
\hline
\end{tabular}
\end{center}
\tablefoot{\tablefoottext{$\dagger$}{The lists include the hfs taken from \cite{Prochaska2000} as well as additional lines nearby. Columns represent element line, wavelength, and log(gf) values.}}

\end{table*}

\begin{table*}[ht]\tiny
\begin{center}
\caption{Line lists for synthesis of the \ion{V}{i} lines.\label{tab:hfs2}\tablefootmark{$^\dagger$}}
\setlength{\tabcolsep}{1 mm}
\begin{tabular}{c c c | c c c  | c c c | c c c | c c c }
\hline
\multicolumn{3}{c|} {\ion{V}{i}  5670  } & \multicolumn{3}{c|}{\ion{V}{i} 6081 } & \multicolumn{3}{c|}{ \ion{V}{i} 6251} & \multicolumn{3}{c|}{ \ion{V}{i} 6274} & \multicolumn{3}{c}{ \ion{V}{i} 6285} \\
Element & $\lambda$ ({\AA}) & log(gf)  & Element & $\lambda$ ({\AA}) & log(gf) & Element & $\lambda$ ({\AA}) & log(gf) & Element & $\lambda$ ({\AA}) & log(gf) & Element &  $\lambda$ ({\AA}) & log(gf)\\
\hline
\ion{Si}{i}  & 5669.736 & -1.780 & \ion{Fe}{i}  & 6078.491 & -0.321 & \ion{Nd}{ii}  &  6250.440 & -1.360 & \ion{Co}{i}  & 6273.004 &  -1.035 & \ion{Co}{i} &  6282.634 & -2.160 \\
\ion{Ni}{i} & 5669.943 & -1.004 & \ion{Fe}{i} & 6079.008 &  -1.120 & \ion{C}{i} & 6250.900 & -0.382 & \ion{Ti}{i} &  6273.388 & -4.008 & \ion{Sm}{ii} & 6284.103 & -1.220\\
\ion{Ti}{i}  & 5670.079 & -1.131& \ion{Si}{i}  & 6080.022 & -2.179 & \ion{Fe}{i}  & 6251.242 & -2.919 & \ion{Fe}{i} & 6274.089 & -1.325 & \ion{Fe} & 6284.528 & -3.317\\
 \ion{V}{i} & 5670.832 & -3.407 &   \ion{V}{i} & 6081.417 & -1.660 &     \ion{V}{i}&  6251.771 & -2.924 &  \ion{V}{i} & 6274.607 & -2.932 &  \ion{V}{i} & 6285.098 & -3.568\\                   
 \ion{V}{i} & 5670.832 & -2.106 &   \ion{V}{i} & 6081.417 & -1.484 &     \ion{V}{i} & 6251.788 & -2.720 &  \ion{V}{i} & 6274.641 & -2.476 &  \ion{V}{i} & 6285.117 & -3.140\\                                   
 \ion{V}{i} & 5670.832 & -1.098 &  \ion{V}{i} & 6081.427 & -1.484 &       \ion{V}{i} & 6251.804 & -2.655 &  \ion{V}{i} & 6274.655 & -2.134 &  \ion{V}{i} & 6285.122 & -2.703\\                                   
 \ion{V}{i} & 5670.844 & -3.009 &   \ion{V}{i} & 6081.428 & -1.359 &      \ion{V}{i} & 6251.806 & -2.021 &  \ion{V}{i} & 6274.657 & -2.455 &  \ion{V}{i} & 6285.134 & -2.890\\                           
 \ion{V}{i} &  5670.844 & -1.896 &  \ion{V}{i} & 6081.442 & -1.359 &      \ion{V}{i} & 6251.817 & -2.662 &  \ion{V}{i} & 6274.678 & -2.601 &   \ion{V}{i} & 6285.137 & -2.542\\                   
 \ion{V}{i} & 5670.844 & -1.203 &  \ion{V}{i} & 6081.442 & -1.677 &       \ion{V}{i} & 6251.818 & -2.208 & \ion{Co}{i}  & 6275.126 & -1.426 & \ion{V}{i} & 6285.148 & -2.714\\
 \ion{V}{i} & 5670.854 & -2.805 &  \ion{V}{i} & 6081.442 & -1.472 &       \ion{V}{i} & 6251.829 & -2.728  & \ion{C}{i} &  6275.555 & -0.673 &  \ion{V}{i} & 6285.149 & -2.550\\
 \ion{V}{i} & 5670.854 & -1.822 &  \ion{V}{i} & 6081.460 & -1.472 &       \ion{V}{i} & 6251.829 & -2.427 & \ion{C}{i} & 6275.576 & -1.345 &  \ion{V}{i} & 6285.152 & -2.077\\
 \ion{V}{i} & 5670.854 & -1.319 & \ion{Fe}{i} & 6081.708 & -2.741 &                               \ion{V}{i} & 6251.837 & -2.690 & & & &  \ion{V}{i} & 6285.157 & -2.714\\
 \ion{V}{i} & 5670.863 & -2.708 &  \ion{Fe}{i} & 6081.829 & -2.852 &                              \ion{V}{i} & 6251.839 & -2.866 & & & &   \ion{V}{i} & 6285.162 & -2.293\\              
 \ion{V}{i} & 5670.863 & -1.816 &  \ion{Co}{i} & 6082.422 & -0.520 &                                     \ion{V}{i} & 6251.844 & -3.021 & & & &  \ion{V}{i} & 6285.168 & -2.576\\
  \ion{V}{i} & 5670.863 & -1.448 & & & &                                          \ion{V}{i} & 6251.848 & -3.146 & &  & &  \ion{V}{i} & 6285.172 & -3.015\\
 \ion{V}{i} & 5670.870 & -2.708 & & & &                                           \ion{V}{i} & 6251.849 & -3.468 & & & &   \ion{Ce}{ii} &  6285.727 & -1.490\\                                                                                   
 \ion{V}{i} & 5670.870 & -1.863 & & & &                                            \ion{V}{i}&  6251.853 & -4.167 & & & &  \ion{Fe}{i} &  6286.509 & -3.447\\                                                                            
 \ion{V}{i} & 5670.870 & -1.594 & & & &                                           \ion{V}{i} & 6251.853 & -2.924 & & & & \ion{V}{i} &  6286.933 &  0.327\\                                                                
 \ion{V}{i} & 5670.875 & -2.863 & & & &                                           \ion{V}{i} & 6251.859 & -3.146 & & &\\                                                                         
 \ion{V}{i} & 5670.875 & -1.968 &  & & &                                          \ion{V}{i} & 6251.859 & -2.720 & & &\\                                                                         
  \ion{V}{i} & 5670.875 & -1.764 & & & &                                  \ion{V}{i} & 6251.862 & -2.655 & & &\\                                                                         
 \ion{V}{i} & 5670.878 & -2.164 & & &&                                     \ion{V}{i} & 6251.863 & -2.866 & & &\\                                                                        
  \ion{V}{i} & 5670.878 & -1.968 & & & &                                          \ion{V}{i} & 6251.864 & -2.728 & & &\\                                                                         
 \ion{V}{i} & 5670.880 & -2.231 & & & &                                           \ion{V}{i} & 6251.864 & -2.662 & & &\\                                                                         
\ion{Cr}{ii} &  5671.657 & -3.806 & & & & \ion{Fe}{i} & 6252.555 & -1.687 & & &\\
\ion{Sc}{i} & 5671.821 & 0.495 & & & & \ion{Fe}{i} & 6253.259 &  -2.793 & & &\\
\ion{Fe}{i} & 5671.829 &  -1.665 & & & & \ion{Si}{i} & 6253.598 & -2.390 & & &\\
\hline
\end{tabular}
\tablefoot{\tablefoottext{$\dagger$}{ The lists include the hfs taken from \cite{Prochaska2000} as well as additional lines nearby. Columns represent element line, wavelength and log(gf) values.}}
\end{center}
\end{table*}

\begin{table*}[ht]\tiny
\begin{center}
\caption{Line lists for synthesis of the \ion{Mn}{i} lines.\label{tab:hfs3}\tablefootmark{$^\dagger$}}
\setlength{\tabcolsep}{1 mm}
\begin{tabular}{c c c | c c c  | c c c | c c c }
\hline
\multicolumn{3}{c|} {\ion{Mn}{i}  5394  } & \multicolumn{3}{c|}{\ion{Mn}{i} 5432 } & \multicolumn{3}{c|}{ \ion{Mn}{i} 6013} & \multicolumn{3}{c}{ \ion{Mn}{i} 6016} \\
Element & $\lambda$ ({\AA}) & log(gf)  & Element & $\lambda$ ({\AA}) & log(gf) & Element & $\lambda$ ({\AA}) & log(gf) & Element &  $\lambda$ ({\AA}) & log(gf) \\
\hline
\ion{Ce}{ii} & 5393.392 & -0.060 & \ion{Fe}{i} &  5431.848 & -2.612 & \ion{Si}{i} & 6012.773 &  -2.130 & \ion{Fe}{i} & 6016.050 &  -2.162\\
\ion{Co}{i} &  5393.740 & -0.604 & \ion{Ti}{i} & 5432.334 & -1.540 & \ion{C}{i} & 6013.170 & -1.314 & \ion{Si}{i} & 6016.210 & -1.645\\
\ion{Fe}{i} &  5394.346 & -2.102 & \ion{Nd}{ii} & 5432.360 &  -1.130 & \ion{C}{i} & 6013.220 &  -1.673 & \ion{C}{i} & 6016.450 & -1.834\\
 \ion{Mn}{i} & 5394.626 & -4.070 &  \ion{Mn}{i} & 5432.506 & -4.377 &  \ion{Mn}{i} & 6013.478 & -0.766 &  \ion{Mn}{i} & 6016.619 & -1.576\\
 \ion{Mn}{i} & 5394.657 & -4.988 &  \ion{Mn}{i} & 5432.510 & -5.155 &  \ion{Mn}{i} & 6013.499 & -0.978 &  \ion{Mn}{i} & 6016.645 & -0.798  \\
 \ion{Mn}{i} & 5394.661 & -4.210 &  \ion{Mn}{i} & 5432.535 & -5.155 &  \ion{Mn}{i} & 6013.518 & -1.251 &  \ion{Mn}{i} & 6016.647 & -1.408\\                                                                
 \ion{Mn}{i} & 5394.684 & -6.205 &  \ion{Mn}{i} & 5432.538 & -4.640 &  \ion{Mn}{i} & 6013.527 & -1.455 &  \ion{Mn}{i} & 6016.667 & -1.408\\
 \ion{Mn}{i} & 5394.687 & -4.812 &  \ion{Mn}{i} & 5432.541 & -4.992 &  \ion{Mn}{i} & 6013.533 & -1.661 &  \ion{Mn}{i} & 6016.668 & -1.061\\                                                
 \ion{Mn}{i} & 5394.690 & -4.368 &  \ion{Mn}{i} & 5432.561 & -4.992 &  \ion{Mn}{i} & 6013.538 & -1.309 &  \ion{Mn}{i} & 6016.684 & -1.392\\
 \ion{Mn}{i} & 5394.709 & -5.853 &  \ion{Mn}{i} & 5432.564 & -4.971 &  \ion{Mn}{i} & 6013.547 & -1.330 &  \ion{Mn}{i} & 6016.685 & -1.510\\
 \ion{Mn}{i} & 5394.712 & -4.786 &  \ion{Mn}{i} & 5432.566 & -4.987 &  \ion{Mn}{i} & 6013.552 & -1.485 &  \ion{Mn}{i} & 6016.696 & -1.839\\
 \ion{Mn}{i} & 5394.714 & -4.552 &  \ion{Mn}{i} & 5432.580 & -4.987 &  \ion{Mn}{i} & 6013.562 & -1.807 &  \ion{Mn}{i} & 6016.696 & -1.576\\
 \ion{Mn}{i} & 5394.728 & -5.728 &  \ion{Mn}{i} & 5432.583 & -5.418 &  \ion{Mn}{i} & 6013.566 & -1.853 &  \ion{Mn}{i} & 6016.698 & -1.772\\
 \ion{Mn}{i} & 5394.730 & -4.853 &  \ion{Mn}{i} & 5432.584 & -5.089 &  \ion{Mn}{i} & 6013.566 & -2.409     &  \ion{Mn}{i} & 6016.704 & -2.538\\
 \ion{Mn}{i} & 5394.731 & -4.774 &  \ion{Mn}{i} & 5432.594 & -5.089 &  \ion{Mn}{i} & 6013.567 & -2.029 &  \ion{Mn}{i} & 6016.707 & -1.413\\
 \ion{Mn}{i} & 5394.741 & -5.807 &  \ion{Mn}{i} & 5432.595 & -6.117 & \ion{Fe}{i} & 6013.915 & -2.133 & \ion{Mn}{i} & 6016.713 & -1.408\\
 \ion{Mn}{i} & 5394.742 & -5.029 &  \ion{Mn}{i} & 5432.596 & -5.351 & \ion{Mn}{i} & 6014.341 & -1.258     &   \ion{Mn}{i} & 6016.714 & -1.772\\   
 \ion{Mn}{i} & 5394.743 & -5.059 &  \ion{Mn}{i} & 5432.601 & -5.351 & \ion{C}{i} & 6014.840 & -1.584 &  \ion{Mn}{i} & 6016.716 & -1.510\\        
 \ion{Fe}{i} & 5395.217 & -2.170 & \ion{Fe}{i} & 5432.948 &  -1.040 & & & & \ion{Pr}{ii} & 6017.800 & -0.257\\
 \ion{Ti}{ii} &  5396.226 & -3.020 & \ion{Fe}{ii} & 5432.967 & -3.527 & & & & \ion{Fe}{i} & 6018.300 & -2.082\\
 \ion{Ti}{ii} & 5396.561 & -1.954 & \ion{Fe}{i} & 5433.116 &  -1.801 & & & & \ion{Ti}{i} & 6018.395 & -1.642\\
\hline
\end{tabular}
\tablefoot{\tablefoottext{$\dagger$}{The lists include the hfs taken from \cite{Prochaska2000} as well as additional lines nearby. Columns represent element line, wavelength and log(gf) values.}}
\end{center}
\end{table*}

\begin{table*}[ht]\tiny
\begin{center}
\caption{Line lists for synthesis of the \ion{Co}{i} lines.\label{tab:hfs4}\tablefootmark{$^\dagger$}}
\setlength{\tabcolsep}{1 mm}
\begin{tabular}{c c c | c c c  | c c c | c c c }
\hline
\multicolumn{3}{c|} {\ion{Co}{i}  5301  } & \multicolumn{3}{c|}{\ion{Co}{i} 5342 } & \multicolumn{3}{c|}{ \ion{Co}{i} 5352} & \multicolumn{3}{c}{ \ion{Co}{i} 5647} \\
Element & $\lambda$ ({\AA}) & log(gf)  & Element & $\lambda$ ({\AA}) & log(gf) & Element & $\lambda$ ({\AA}) & log(gf) & Element & $\lambda$ ({\AA}) & log(gf) \\
\hline
\ion{Cr}{i} &  5300.745 &  -2.000 & \ion{Sc}{ii} & 5342.050 & -3.053 & \ion{Ti}{i} & 5351.068 & 0.010 & \ion{Fe}{i} & 5645.833 & -0.961\\
\ion{Cr}{i} & 5300.746 & -1.822 & \ion{Fe}{i} & 5342.403 & -2.116 & \ion{Co}{i} & 5351.407 &  -0.732 & \ion{V }{i} & 5646.108 & -1.190\\
\ion{Fe}{i} & 5300.864 & -2.930 & \ion{Fe}{i} & 5342.545 & -1.901 & \ion{Ni}{i} & 5351.839 &  -2.514 & \ion{Fe}{i} & 5646.684 & -2.500\\
 \ion{Co}{i} & 5301.008 & -3.345 &  \ion{Co}{i} & 5342.693 & -0.640 &  \ion{Co}{i} & 5351.976 & -1.573 &  \ion{Co}{i} & 5647.207 & -2.127 \\
 \ion{Co}{i} & 5301.017 & -2.646 &  \ion{Co}{i} & 5342.693 & -0.557 &  \ion{Co}{i} & 5351.978 & -2.584 &  \ion{Co}{i} & 5647.220 & -2.343 \\
 \ion{Co}{i} & 5301.026 & -3.164 &  \ion{Co}{i} & 5342.693 & -0.471 &  \ion{Co}{i} & 5351.997 & -1.491 &  \ion{Co}{i} & 5647.232 & -2.626 \\
 \ion{Co}{i} & 5301.034 & -2.937 &  \ion{Co}{i} & 5342.694 & -0.387 &  \ion{Co}{i} & 5352.000 & -2.612 &  \ion{Co}{i} & 5647.239 & -2.753 \\
 \ion{Co}{i} & 5301.041 & -3.133 &  \ion{Co}{i} & 5342.696 & -0.305 &  \ion{Co}{i} & 5352.015 & -1.475 &  \ion{Co}{i} & 5647.243 & -3.065\\
 \ion{Co}{i} & 5301.043 & -3.345 &  \ion{Co}{i} & 5342.699 & -0.228 &  \ion{Co}{i} & 5352.019 & -2.818 &  \ion{Co}{i} & 5647.246 & -2.592\\
 \ion{Co}{i} & 5301.048 & -3.344 &  \ion{Co}{i} & 5342.701 & -1.365 &  \ion{Co}{i} & 5352.020 & -0.644 &  \ion{Co}{i} & 5647.253 & -2.600\\
 \ion{Co}{i} & 5301.052 & -3.197 &  \ion{Co}{i} & 5342.702 & -0.154 &  \ion{Co}{i} & 5352.029 & -1.511 &  \ion{Co}{i} & 5647.259 & -2.764\\
 \ion{Co}{i} & 5301.056 & -3.164 &  \ion{Co}{i} & 5342.704 & -1.147 &  \ion{Co}{i} & 5352.036 & -0.728 &  \ion{Co}{i} & 5647.265 & -3.618\\
 \ion{Co}{i} & 5301.058 & -4.061 &  \ion{Co}{i} & 5342.706 & -0.084 &  \ion{Co}{i} & 5352.041 & -1.607 &  \ion{Co}{i} & 5647.267 & -3.190\\
 \ion{Co}{i} & 5301.061 & -3.396 &  \ion{Co}{i} & 5342.707 & -1.053 &  \ion{Co}{i} & 5352.049 & -0.818 &  \ion{Co}{i} & 5647.269 & -2.764\\
 \ion{Co}{i} & 5301.065 & -5.305 &  \ion{Co}{i} & 5342.712 & -1.021 &  \ion{Co}{i} & 5352.050 & -1.818 &  \ion{Co}{i} & 5647.269 & -2.940\\
 \ion{Co}{i} & 5301.065 & -3.133 &  \ion{Co}{i} & 5342.713 & -2.578 &  \ion{Co}{i} & 5352.059 & -0.914 & \ion{Fe}{i} & 5647.772 & -4.943\\
 \ion{Co}{i} & 5301.070 & -3.651 &  \ion{Co}{i} & 5342.716 & -1.041 &  \ion{Co}{i} & 5352.066 & -1.018 & \ion{Cr}{i} & 5647.893 & -0.909\\
 \ion{Co}{i} & 5301.071 & -3.197 &  \ion{Co}{i} & 5342.718 & -2.344 &  \ion{Co}{i} & 5352.070 & -1.366 & \ion{Cr}{i} & 5648.262 & -1.000\\
 \ion{Co}{i} & 5301.074 & -3.396 &  \ion{Co}{i} & 5342.722 & -1.128 &  \ion{Co}{i} & 5352.070 & -1.130 & &\\
\ion{Fe}{i} & 5301.313 & -2.750&  \ion{Co}{i} & 5342.725 & -2.298 &  \ion{Co}{i} & 5352.072 & -1.248 & & \\
\ion{Fe}{i} & 5301.862 & -2.650 &  \ion{Co}{i} & 5342.728 & -1.351 & \ion{Pr}{ii} & 5352.404 &  -0.653&\\
\ion{La}{ii} & 5301.970 & -1.140 &  \ion{Co}{i} & 5342.732 & -2.365 & \ion{Yb}{ii} & 5352.954 & -0.340 &\\
& & &  \ion{Co}{i} & 5342.739 & -2.548 & \ion{Fe}{i} & 5353.374 &  -0.840 &\\
& & &  \ion{Co}{i} & 5342.748 & -2.930 & & & &\\
& & & \ion{Ti}{i} & 5342.912 & -1.587 & & & & \\
& & & \ion{Sc}{i} &  5342.958 & -2.275 & & & & \\
& & & \ion{Co}{i} & 5343.380 &  -0.226 & & & &\\
\hline
\end{tabular}
\tablefoot{\tablefoottext{$\dagger$}{ The lists include the hfs taken from \cite{Prochaska2000} as well as additional lines nearby. Columns represent element line, wavelength and log(gf) values.}}
\end{center}
\end{table*}

\section{Solar abundance plots}

\begin{figure*}[ht]
\centering
\resizebox{\hsize}{!}{
\includegraphics[bb=0 50 792 612,clip]{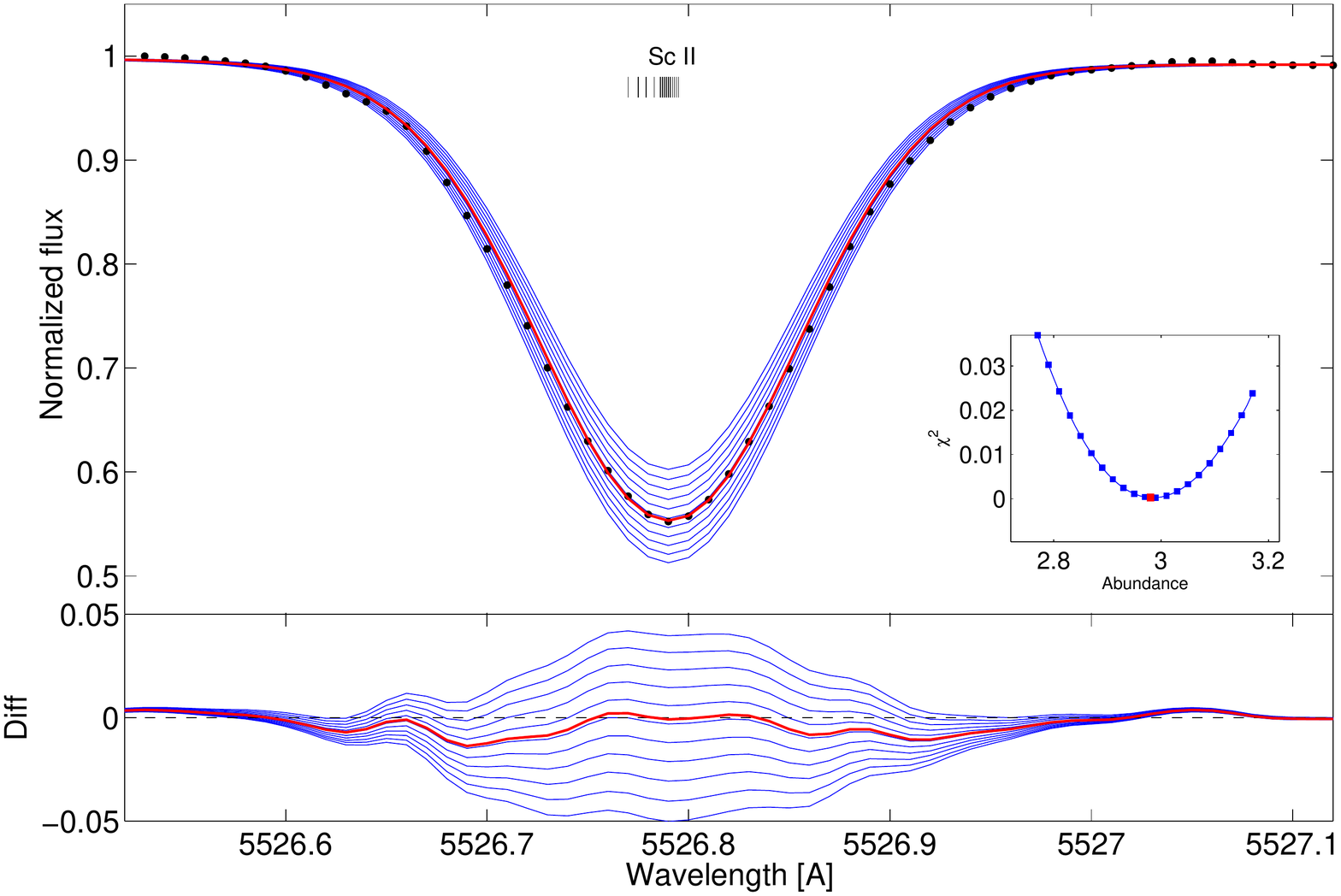} 
\includegraphics[bb=0 50 792 612,clip]{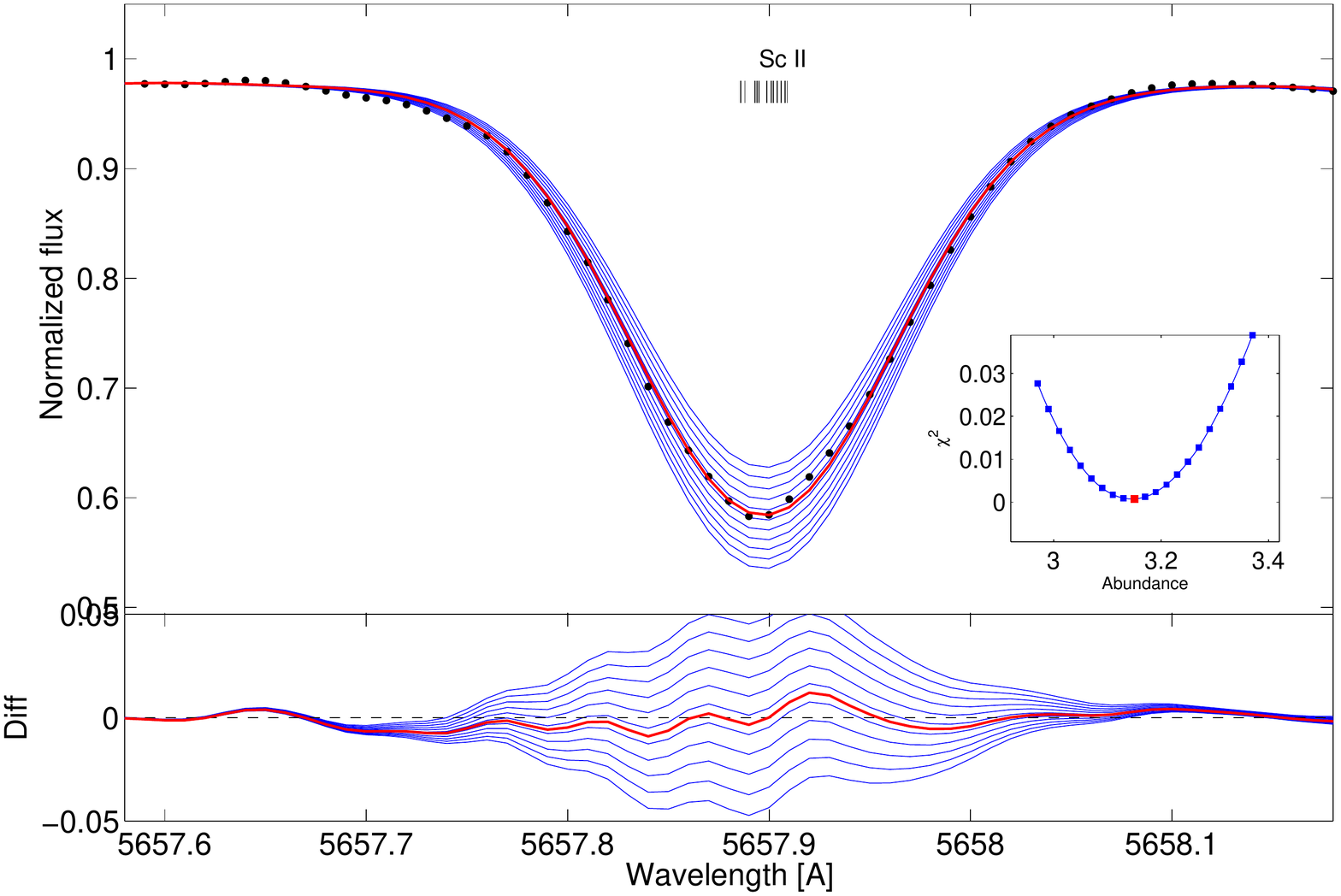} }
\resizebox{\hsize}{!}{
\includegraphics[bb=0 50 792 612,clip]{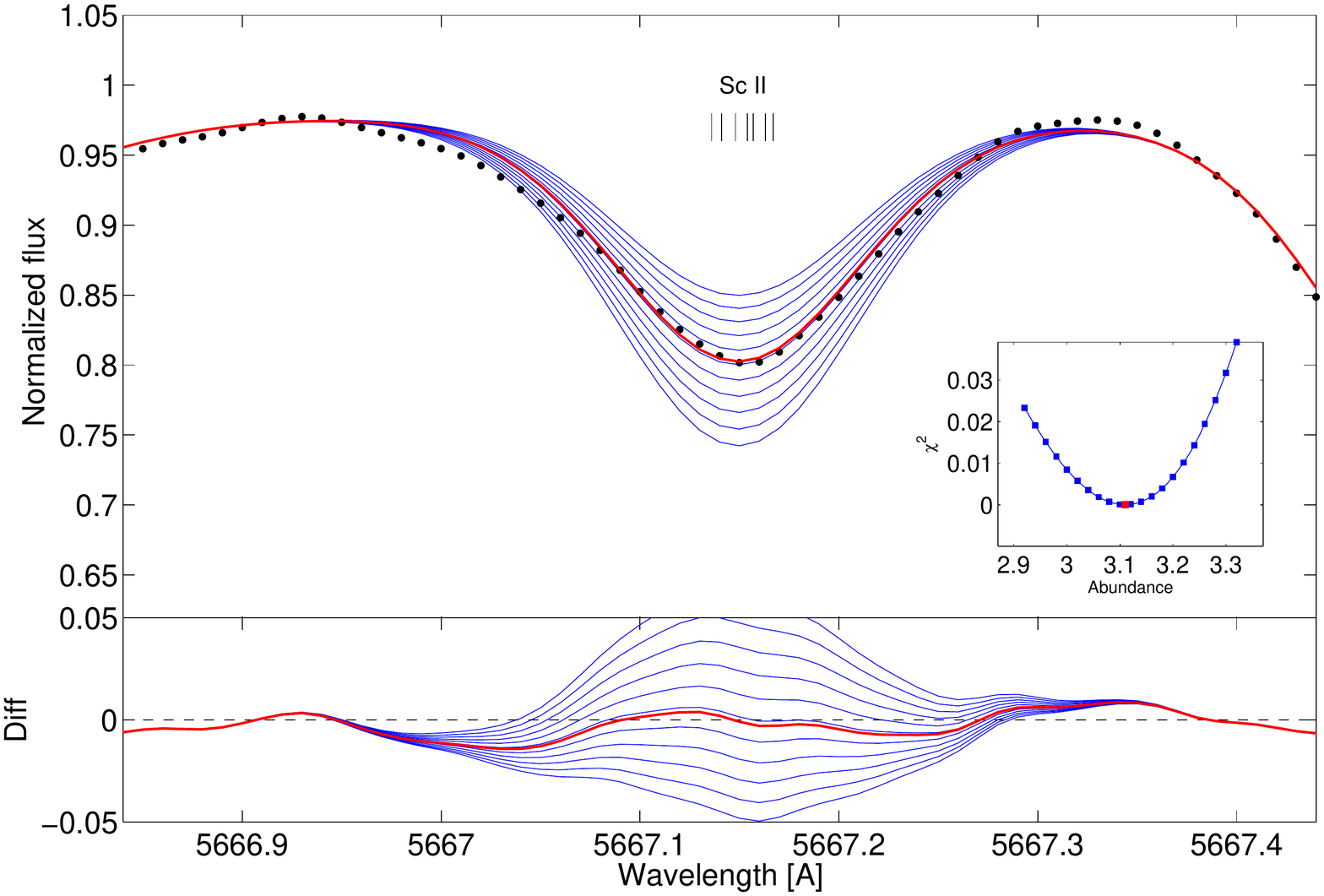} 
\includegraphics[bb=0 50 792 612,clip]{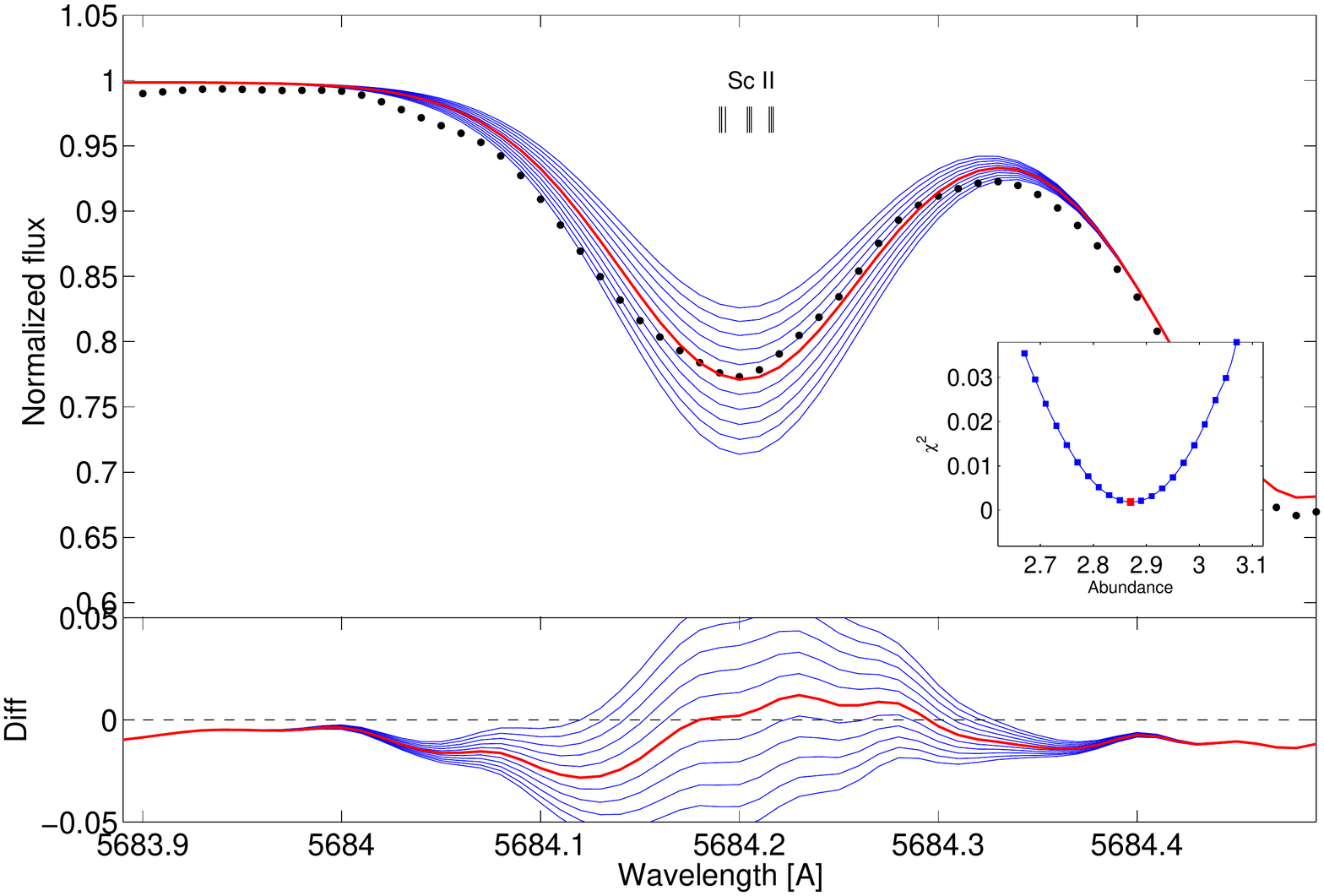} }
\resizebox{\hsize}{!}{
\includegraphics[bb=-396 50 1188 612,clip]{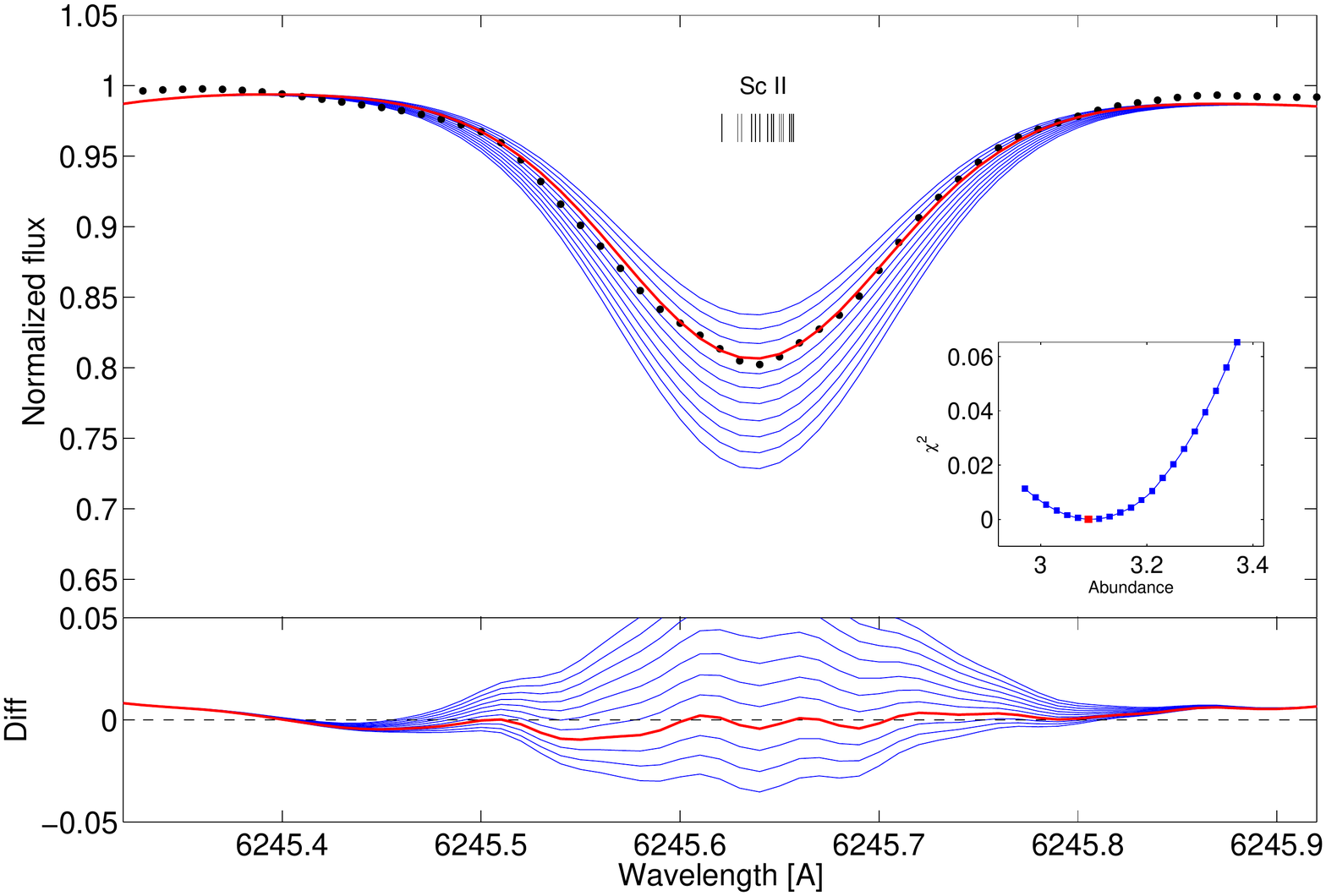}}
\caption{Solar spectrum taken using asteroid Vesta during run at Magellan in January 2006. Spectral lines are listed in order of element and wavelength. Chemical elements with all the hfs lines are indicated inside the lines. The different colored lines are the different synthetic spectra with different abundances in steps of 0.04\,dex, while the dots are the observed spectra. In the lower panels are the values of differences between the real and synthetic spectra for the synthetic spectra plotted above. The red lines represent the best fit derived from unnormalized $\chi^{2}$,   visible in the small plot as a red dot. Here the fits for the five Sc lines are shown.\label{fig:solar_spectrum1} }
\end{figure*}

\begin{figure*}[!ht]
\resizebox{\hsize}{!}{
\includegraphics[bb=0 50 792 612,clip]{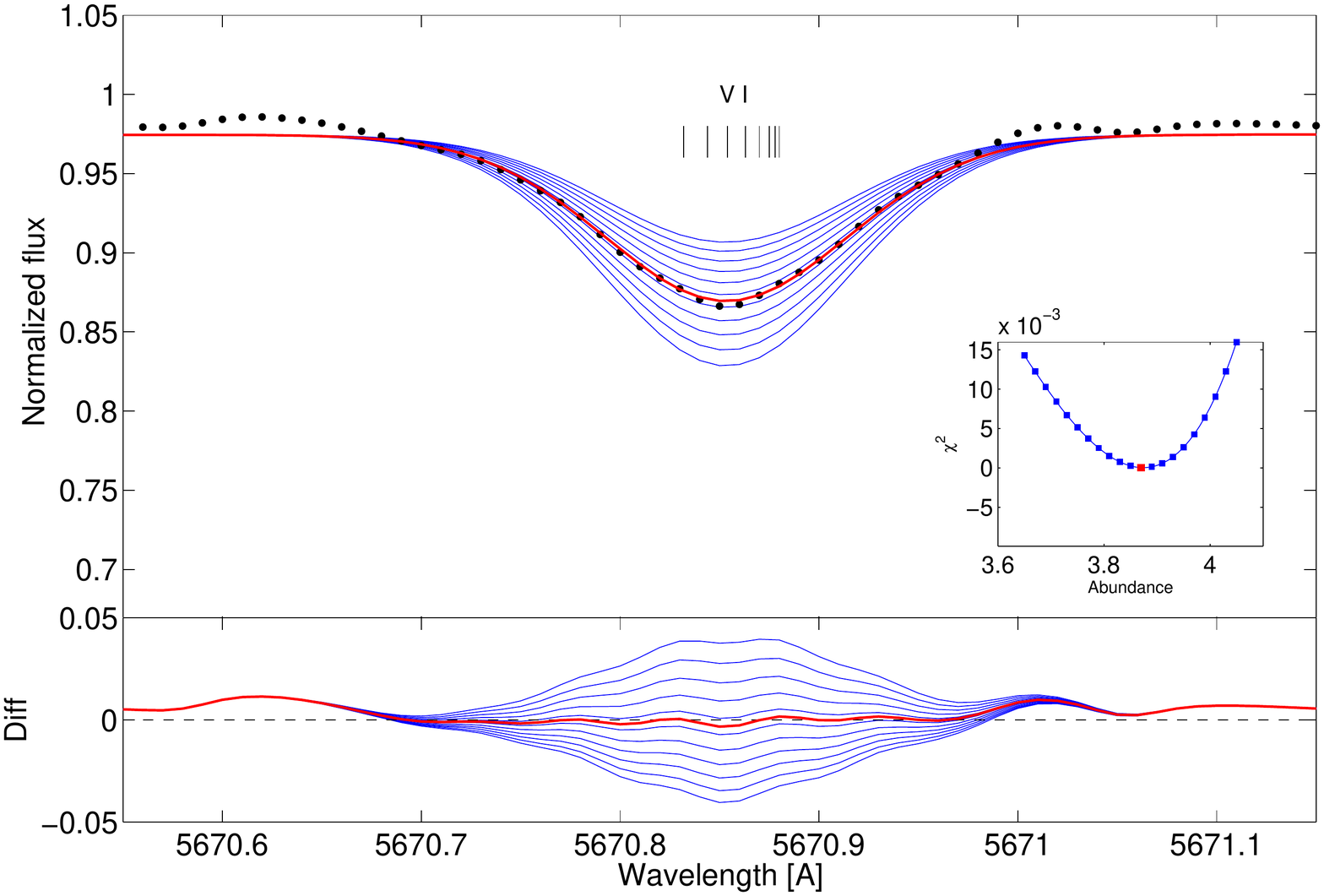}
\includegraphics[bb=0 50 792 612,clip]{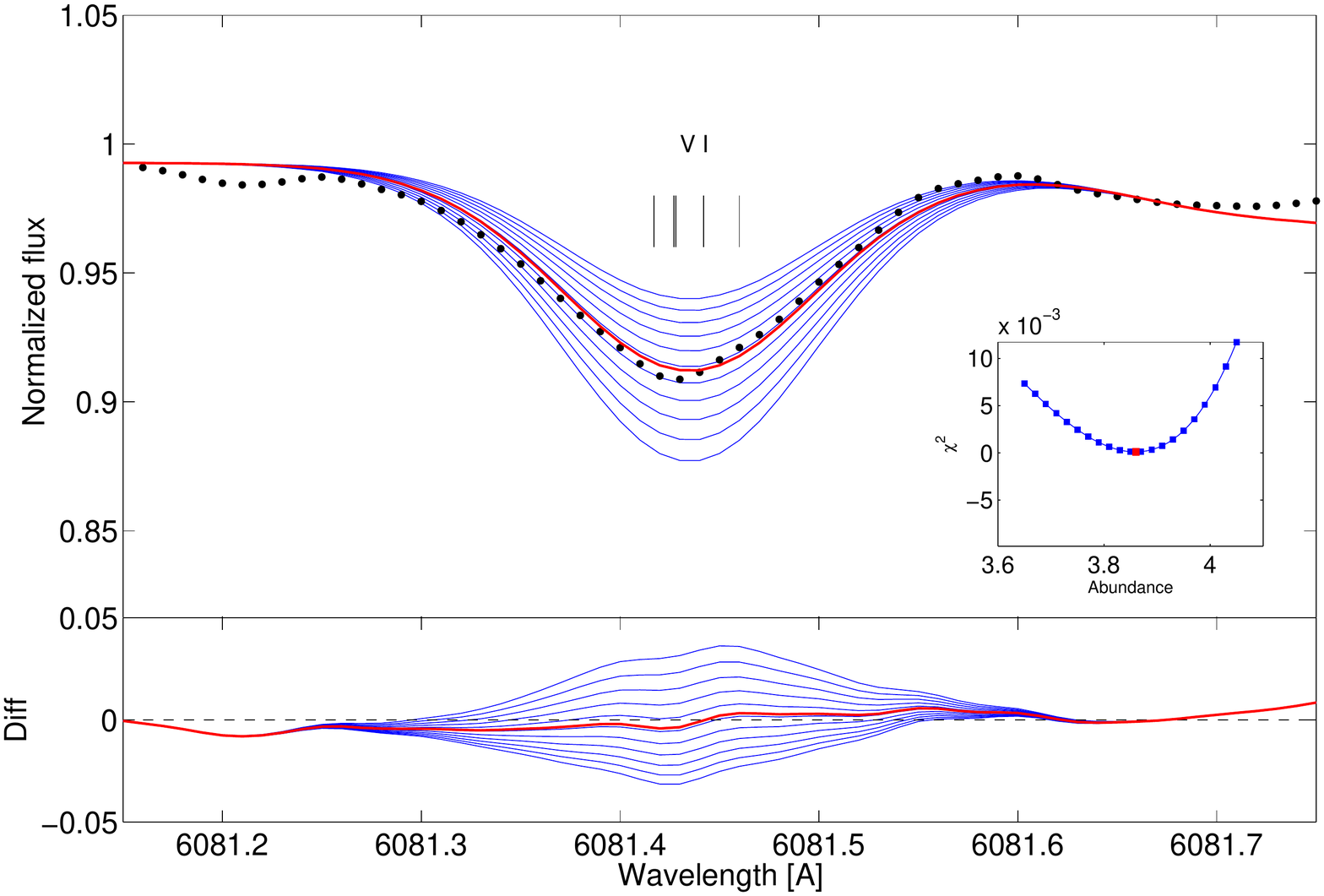}}
\resizebox{\hsize}{!}{
\includegraphics[bb=0 50 792 612,clip]{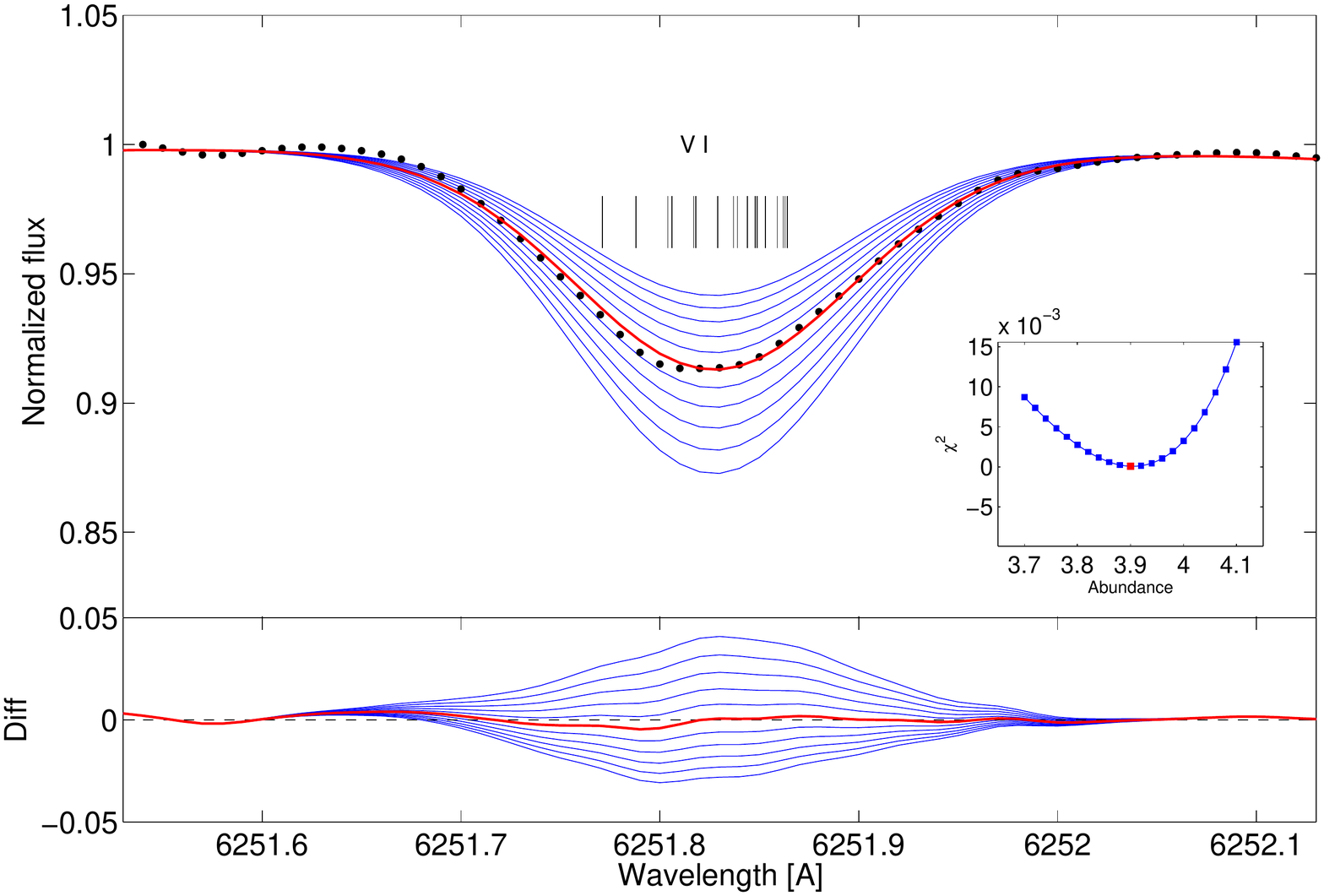}
\includegraphics[bb=0 50 792 612,clip]{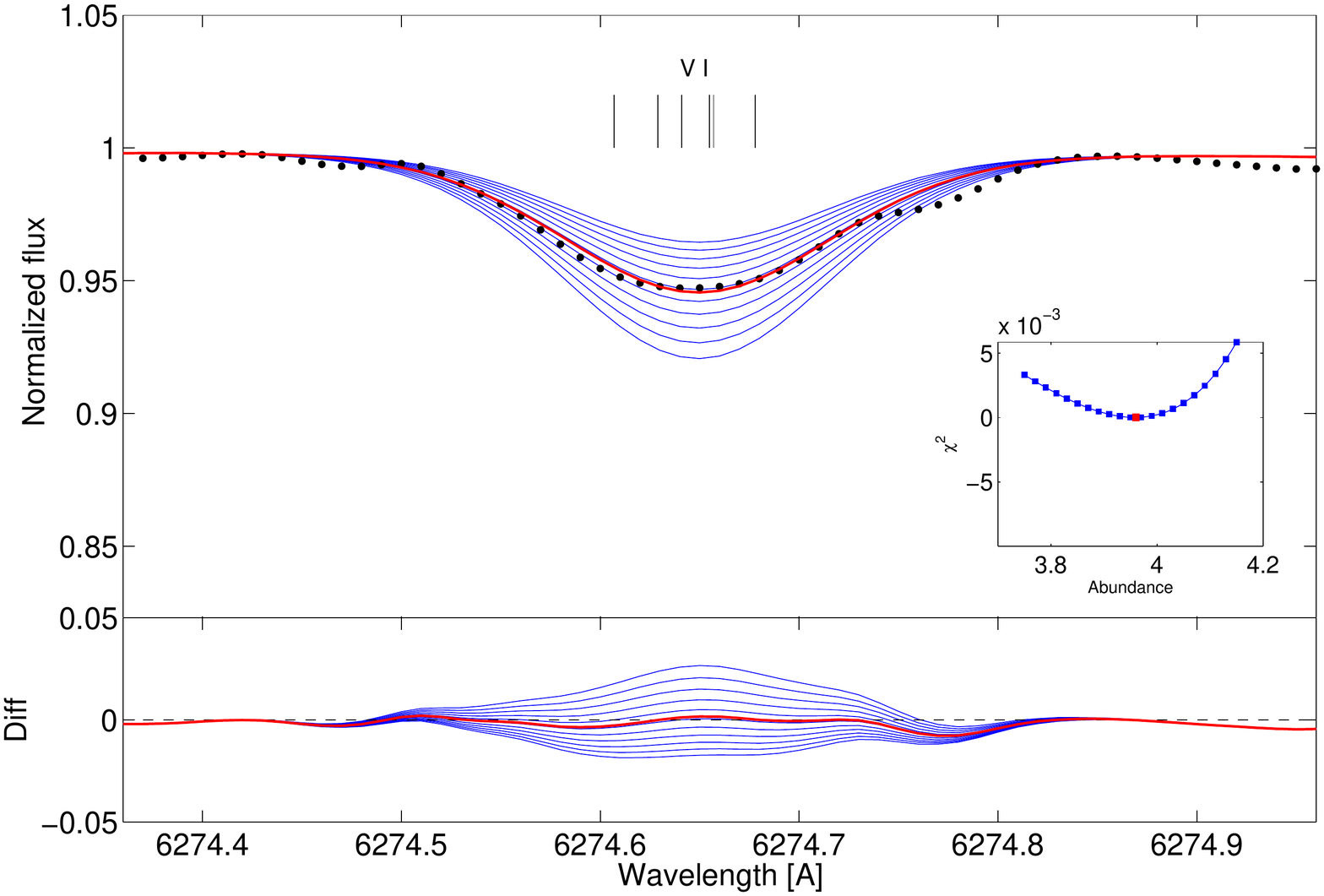}}
\resizebox{\hsize}{!}{
\includegraphics[bb=-396 50 1188 612,clip]{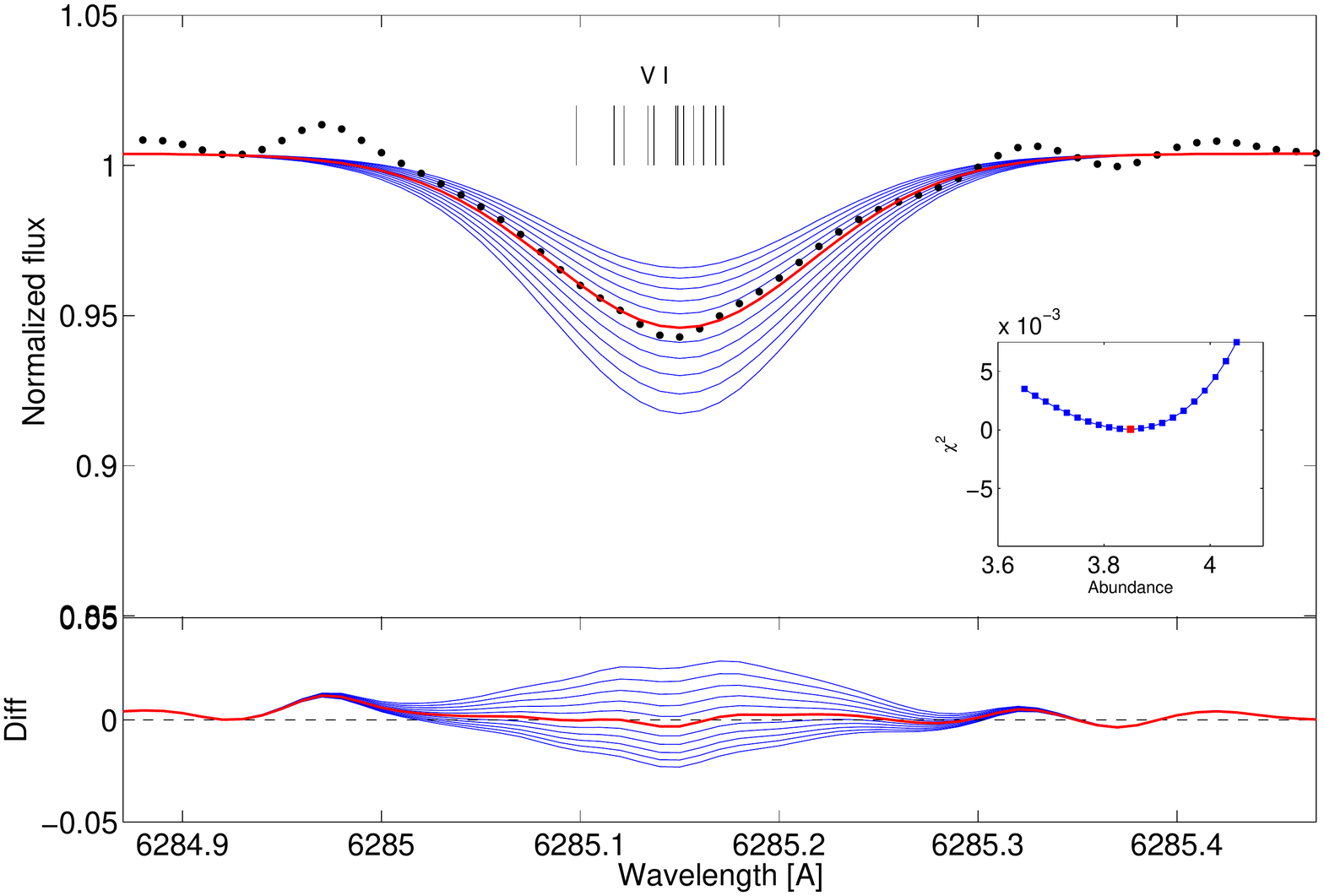}}
\caption{As in Fig.~\ref{fig:solar_spectrum1} but for the five V lines. \label{fig:solar_spectrum2}}
\end{figure*}

\begin{figure*}[!ht]
\resizebox{\hsize}{!}{
\includegraphics[bb=0 50 792 612,clip]{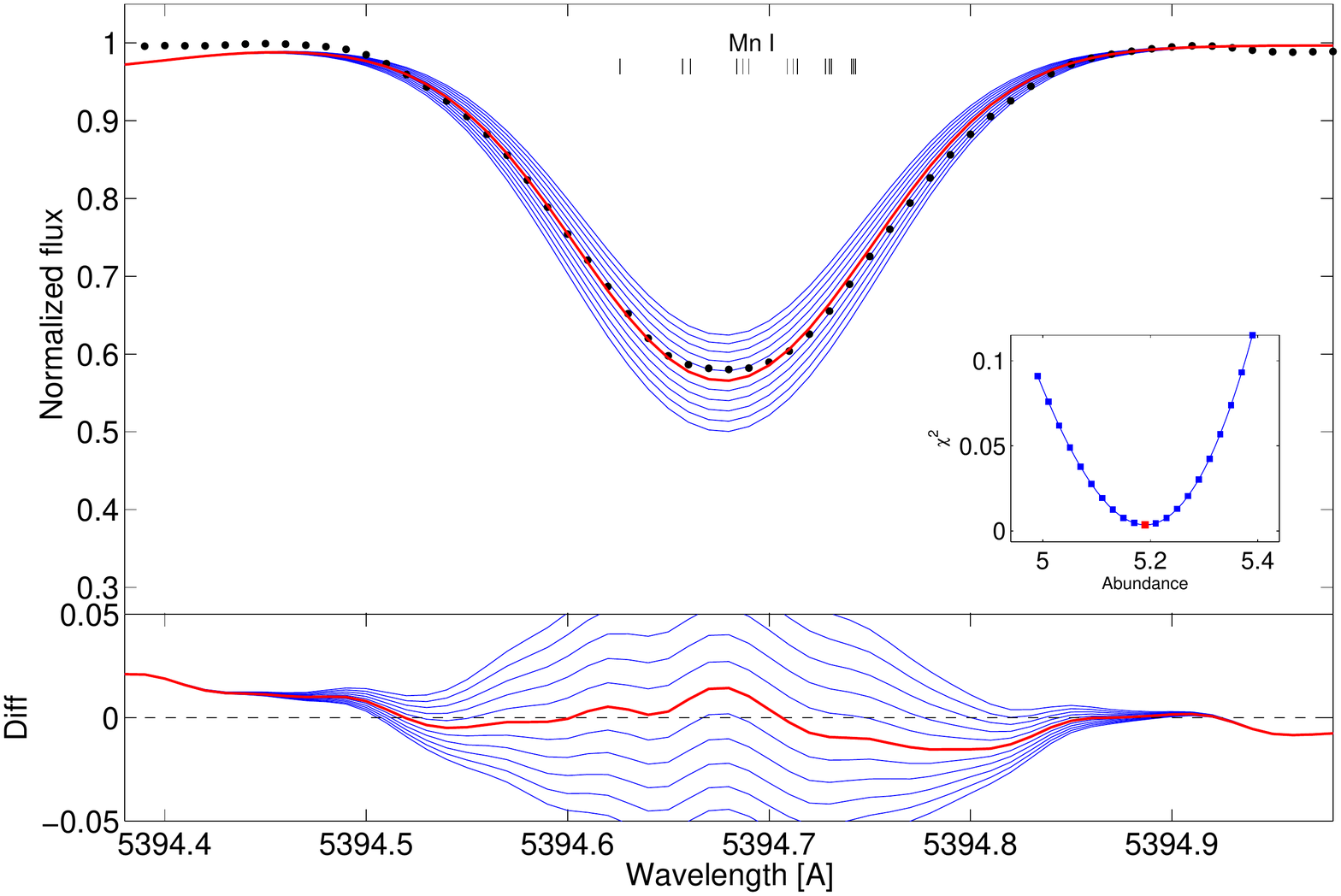}
\includegraphics[bb=0 50 792 612,clip]{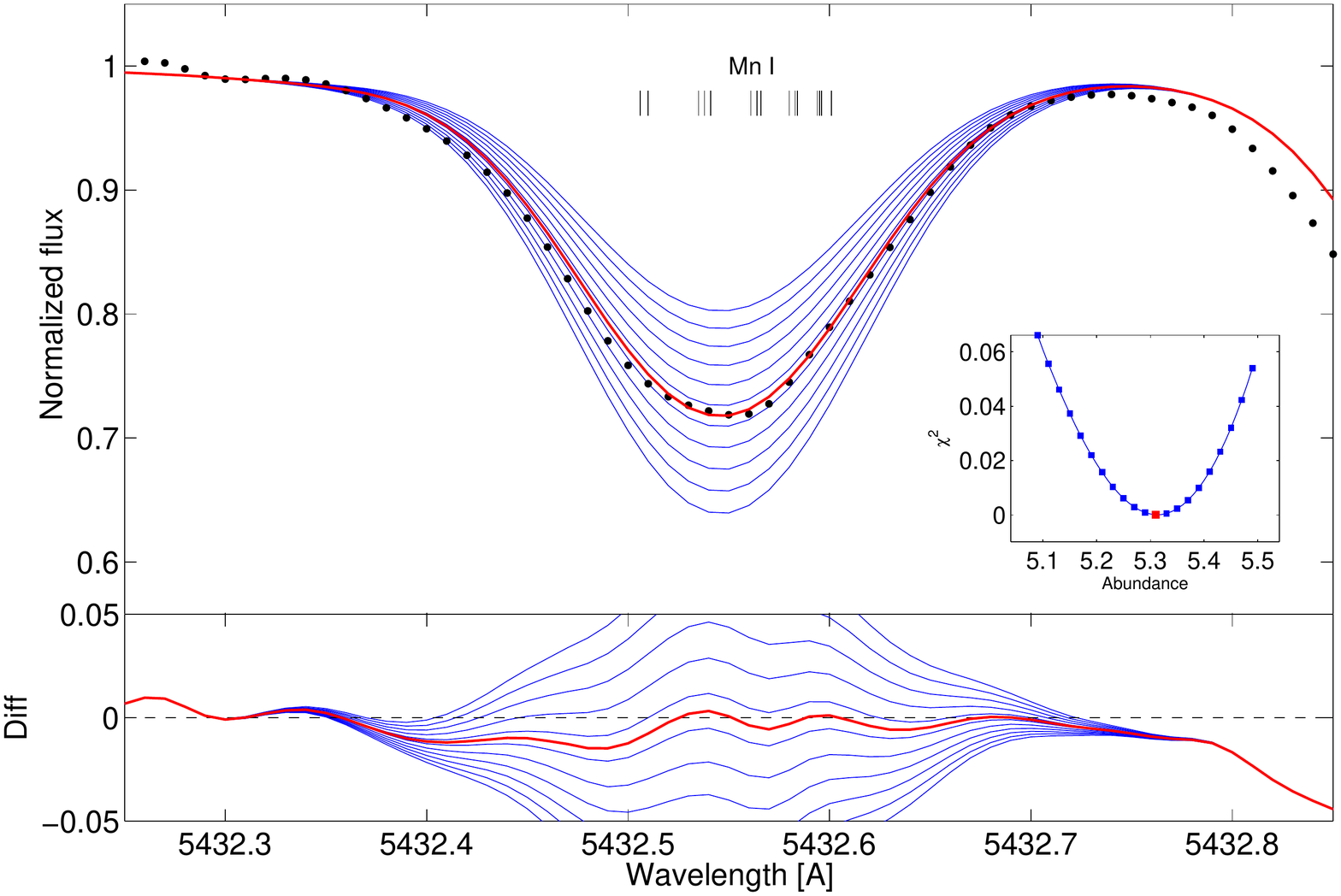}}
\resizebox{\hsize}{!}{
\includegraphics[bb=0 50 792 612,clip]{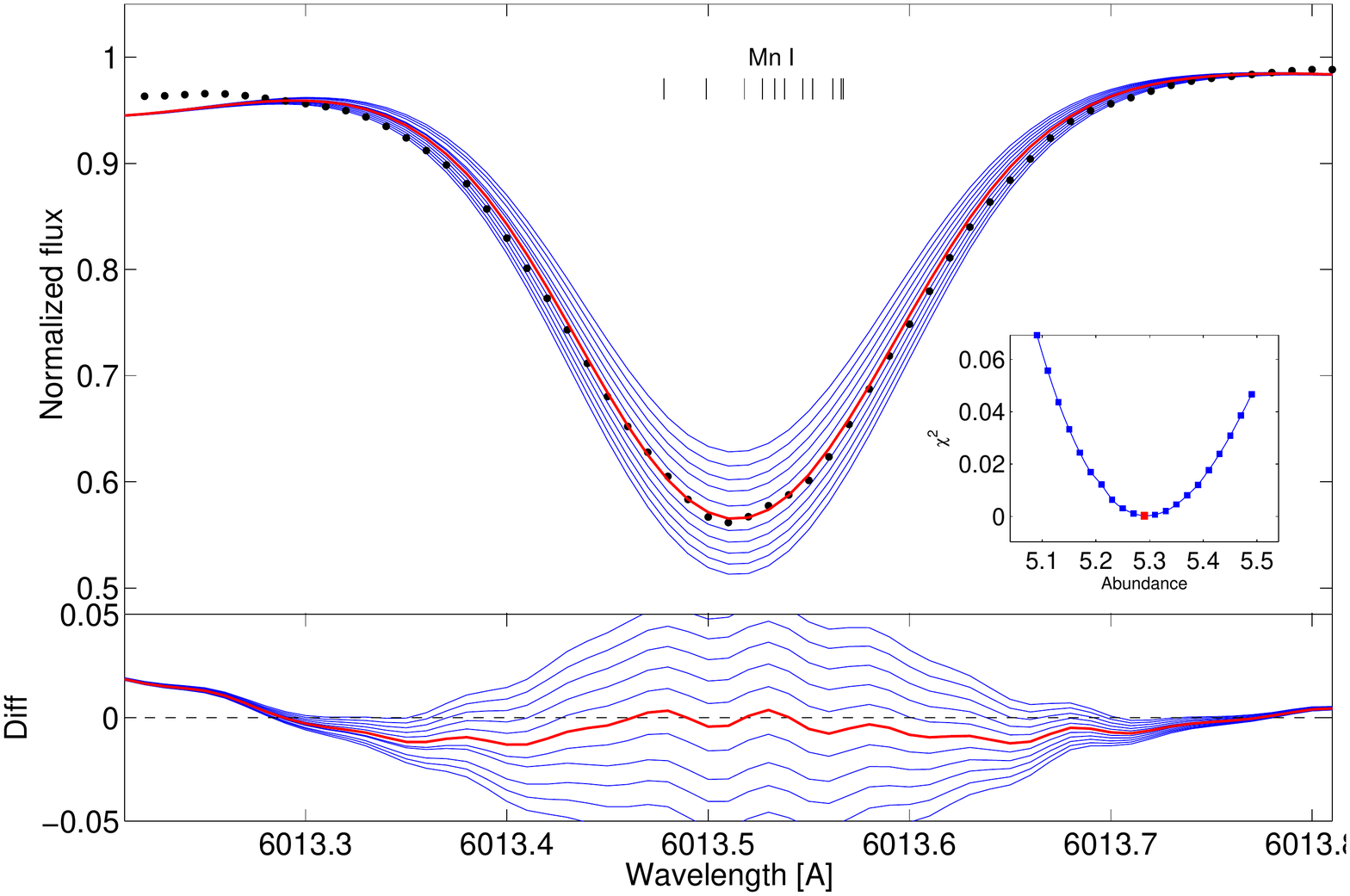}
\includegraphics[bb=0 50 792 612,clip]{Mn6016_2_ab_hipVesta_456.pdf}}
\caption{As in Fig.~\ref{fig:solar_spectrum1} but for the four Mn lines. \label{fig:solar_spectrum3}}
\end{figure*}

In Figs.~\ref{fig:solar_spectrum1}--\ref{fig:solar_spectrum3}  the synthesis for all the lines analyzed in this work for  the solar spectrum from Vesta observed at Magellan in January 2006 is shown. Each plot shows (in the bigger panel) the line fitting with the best fit  to the observed one  and the different abundances in steps of 0.04\,dex. In the lower panel the differences between the observed spectrum and the synthetic one are shown, with the difference with the best fit highlighted. In the small square panel  the $\chi^{2}$ values for the different abundances are represented, with the red dot representing the minimum and therefore the best fit value for the elemental abundance.

\begin{figure*}[!ht]
\resizebox{\hsize}{!}{
\includegraphics[bb=0 50 792 612,clip]{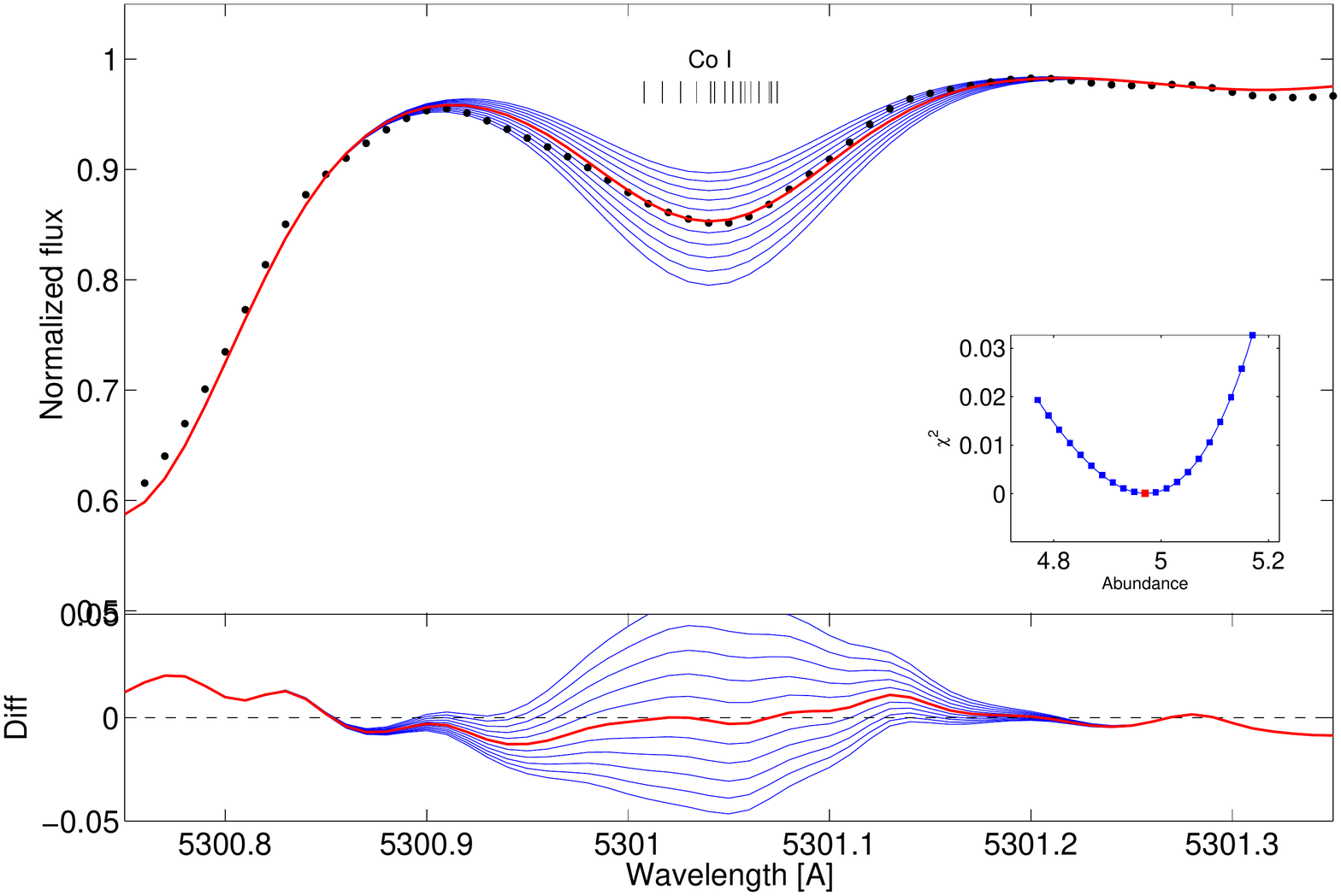}
\includegraphics[bb=0 50 792 612,clip]{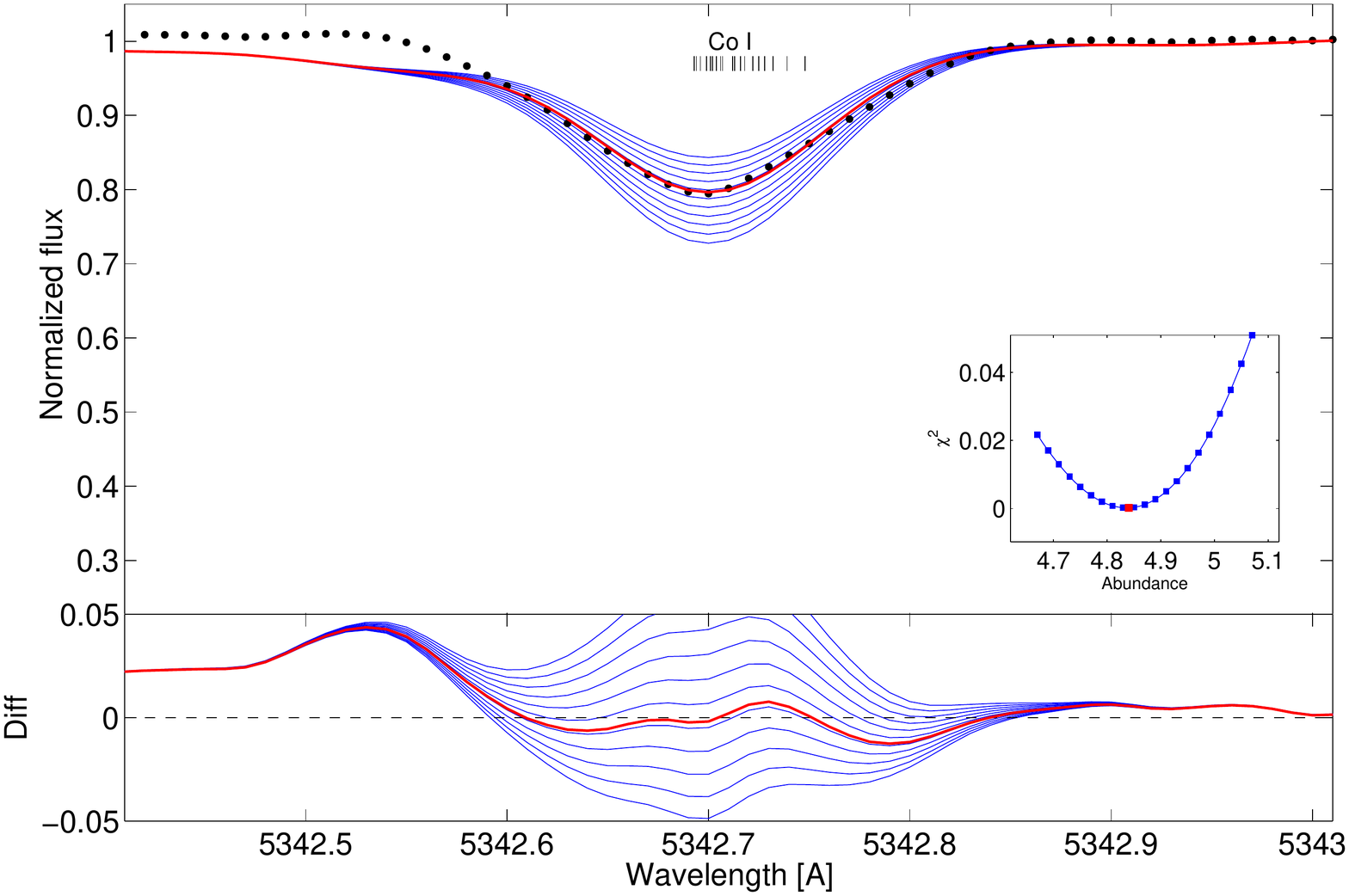}}
\resizebox{\hsize}{!}{
\includegraphics[bb=0 50 792 612,clip]{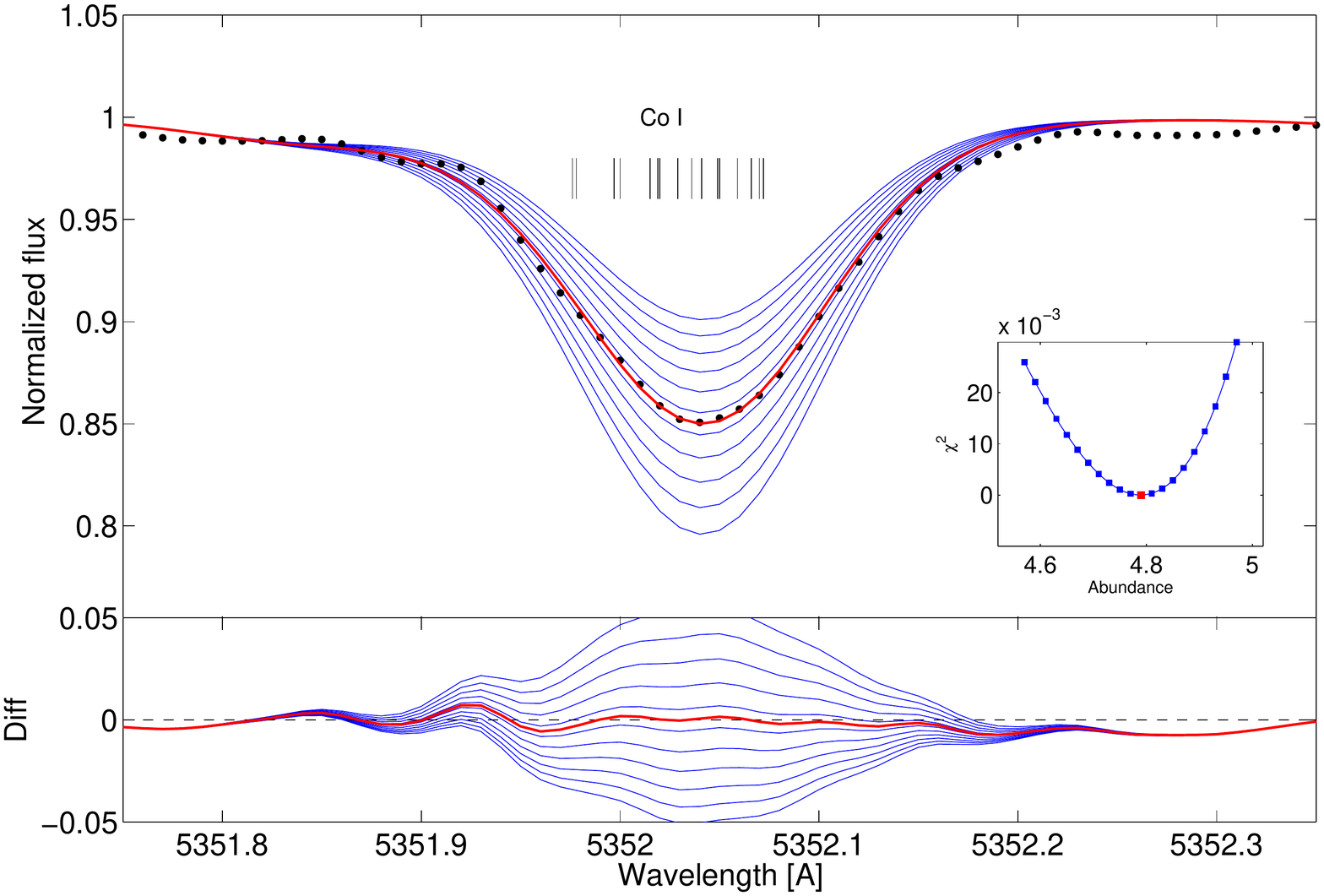}
\includegraphics[bb=0 50 792 612,clip]{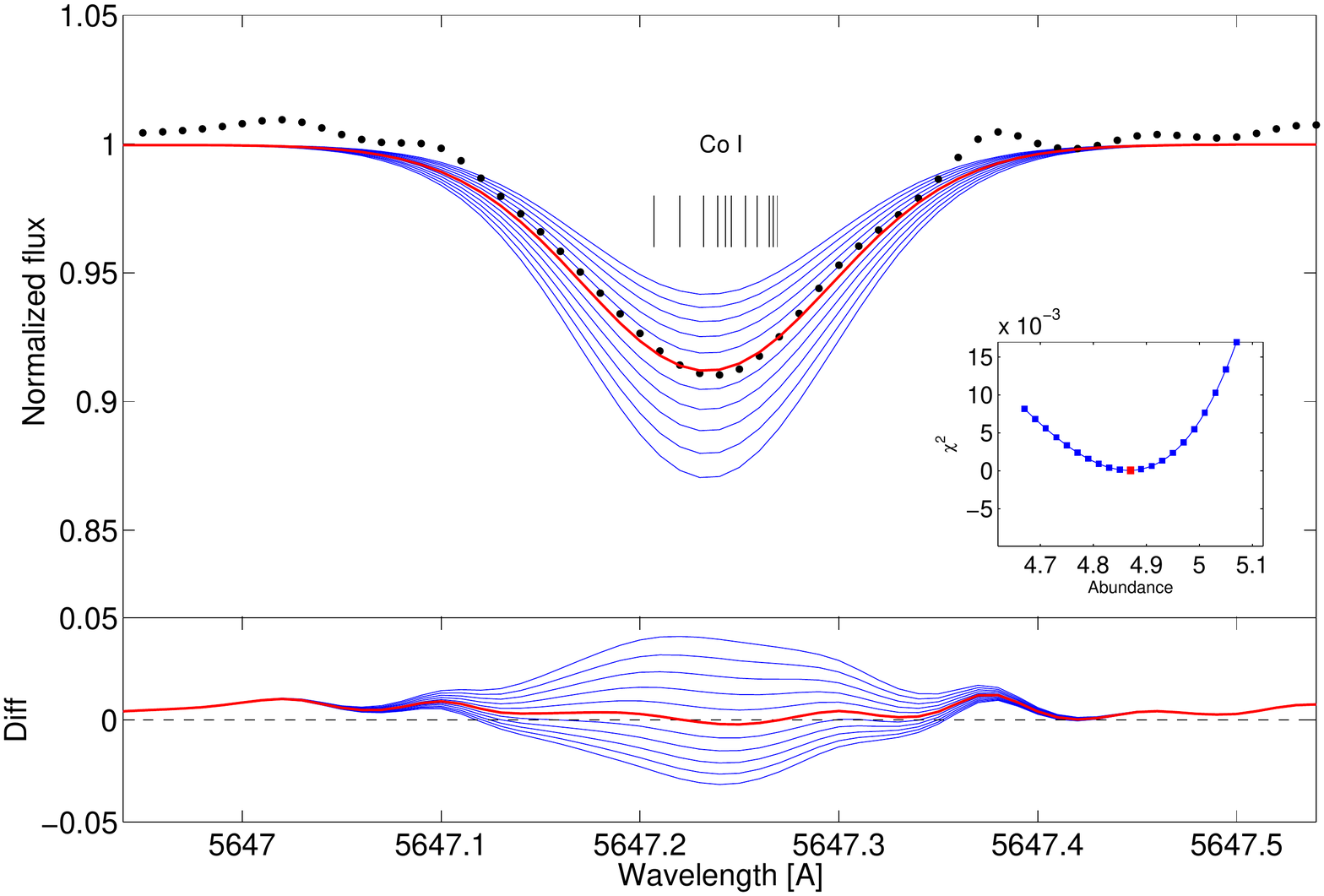}}
\caption{As in Fig.~\ref{fig:solar_spectrum1} but for the four Co lines. \label{fig:solar_spectrum4}}
\end{figure*}

\end{appendix}

\end{document}